\documentclass[%
 preprint,
preprint,
 amsmath,amssymb,
 aps,
 pra,
floatfix,
]{revtex4-2}

\usepackage{ulem}
\usepackage{graphicx}% Include figure files
\usepackage{subfigure}% Include subfigure structures
\usepackage{xcolor}% Include for tex colored
\usepackage{dcolumn}% Align table columns on decimal point
\usepackage{bm}% bold math
\usepackage{hyperref}% add hypertext capabilities
\begin{document}

\preprint{PRE/v4}

\title{Mean-Field Theory for the Three-State Active Lattice Gas Model}%Force line breaks with //

\author{Ana L. N. Dias}
\email{anand@ufmg.br}
\affiliation{%
 Departamento de Física, ICEx, Universidade Federal de Minas Gerais, C. P. 702, 30123-970 Belo Horizonte, Minas Gerais, Brazil.
}
\author{Tiago Venzel Rosembach}
\email{tiagovenzel@cefetmg.br}
\affiliation{Departamento de Formação Geral do {\it campus} Leopoldina, Centro Federal de Educação Tecnológica de Minas Gerais, Rua José Peres, 558 - Centro - Leopoldina-MG, 36700-001, Brazil.}
\author{Ronald Dickman}
\email{dickman@fisica.ufmg.br}
\affiliation{%
 Departamento de Física, ICEx, Universidade Federal de Minas Gerais, C. P. 702, 30123-970 Belo Horizonte, Minas Gerais, Brazil.
}

%\date{\today}

\begin{abstract}
We develop a mean-field description including spatial structure for a simplified version of the three-state active matter model studied by Venzel \textit{et al.} (Phys. Rev. E \textbf{110}, 014109 (2024)). The resulting triangular lattice of coupled nonlinear differential equations are integrated numerically using a fourth-order Runge-Kutta scheme. Starting from various ordered initial configurations, we probe the stability of the corresponding stationary states, revealing the presence of various high-density ordered structures in the density-noise plane. The results are compared with Monte Carlo simulations of the simplified model, yielding, in certain cases, unexpected transitions between ordered configuration types. 
\end{abstract}

\keywords{active matter; active lattice gas; mean field theory}

\maketitle
\newpage

%\tableofcontents

\section{\label{sec:int} Introduction}

\textit{Active matter} refers to systems composed of entities that absorb and dissipate energy to generate their own motion. These systems are inherently out of equilibrium, as they continuously consume energy to sustain activity, and may lead to the emergence of collective behaviors arising from interactions among the self-propelled agents and their environment. Active matter spans a range of scales and fields, intersecting nonequilibrium statistical physics, soft matter, computational modeling, and biology. Its applications extend from understanding biological collectives, like flocks of birds \cite{ballerini2008} and sheep herding \cite{ginelli2015}, to exploring novel behaviors in cell motility \cite{volker2010} and bacterial colonies \cite{zhang2010}.

A landmark in the study of active matter systems is the model introduced by Vicsek \textit{et al.} \cite{vicsek1995}, aimed at understanding how systems of particles with orientational interactions are capable of organizing into collective motion without a leader. In the Vicsek model (VM), particles are subject to alignment interactions with their neighbors and undergo a phase transition from a disordered phase -- with randomly oriented velocities -- to an ordered, \textit{flocking} phase in which particles align along a spontaneously chosen direction. The control parameters of this transition are the density $\rho_0$, and the intensity $\eta$ of the noise representing fluctuations in the direction of motion of particles. Near the critical point, microphase separation occurs, characterized by the coexistence of ordered and disordered regions. This phase separation indicates that the transition in the Vicsek model is discontinuous, a feature that becomes apparent only in large systems \cite{chate2004}. This highlights the impact of finite-size effects in active matter models, since the fluctuations that desestabilize homogeneity near the critical point -- and drive the phase separation -- are suppressed in small systems.

The flocking phase and phase coexistence are also observed in active matter systems in discrete spaces. In the active Ising model (AIM) \cite{solon2013, solon2015b}, the flocking phase is preceded by the emergence of a transverse liquid band that moves through a disordered gaseous background. In the active Potts model (APM) with $q=4$ and $q=6$ states \cite{chatterjee2020, mangeat2020}, phase coexistence can involve a transverse band or a longitudinal band, depending on the particle speed and temperature. In these models, the phase transition is interpreted as a liquid-gas transition, where the liquid phase represents ordered collective motion and the gas corresponds to the disordered state. It was later shown that the transition to the flocking phase in the VM can also be understood as a liquid-gas transition \cite{solon2015c}. The difference is that the VM exhibits microphase separation, whereas AIM and APM display macrophase separation. The active clock model with $q$ orientations is a discretized version of the Vicsek model \cite{karmakar2024}; for small $q$, it behaves similarly to the APM \cite{mangeat2020}, while for $q \to \infty$, it reproduces the phenomenology of the original VM \cite{karmakar2024, chatterjee2022, solon2022}.

Of particular interest are models that incorporate particle repulsion. In discrete spaces, this is readily implemented through excluded-volume interactions that restrict site occupancy. Notably, no lattice model with excluded-volume interactions has been shown to exhibit a flocking phase so far. Nevertheless, the excluded-volume interaction gives rise to unique organizational patterns due to particle congestion in high-density regions \cite{karmakar2022}. In the APM with $q=4$ states proposed by Peruani \textit{et al.} \cite{peruani2011}, ordered structures such as traffic jams, gliders (clusters with two orientantions moving across the lattice), and immobile bands coexisting with a disordered phase have been reported. These structures arise due to the combined effects of excluded volume and local alignment, the strength of the latter being controlled by a parameter $g \ge 0$.

The three-state active lattice gas model (3-SALGM) \cite{rosembach2024} motivates the research developed in this work. This model is defined on a triangular lattice with only three possible velocity orientations, featuring an alignment interaction with nineteen neighbors (including the particle itself), a noise term $\eta$ that disrupts this alignment, and excluded-volume interactions. Simulations show that the ordered phase is characterized by the coexistence of a dense, ordered structure with low mobility and a low-density vapor phase. Depending on the control parameters -- density and noise -- as well as the initial conditions and system size, the three-state active lattice gas can exhibit immobile bands, mobile bands, and traffic jams.

The aim of the present work is to develop a mean-field theory (MFT) for the active lattice gas model, incorporating spatial structure that permits investigation of the robustness of different condensed structures. For this, equations of motion are constructed that govern the probability of finding a particle in each orientation at each site over time, considering that the state of an active particle is subject to changes due to the influence of particles in its neighborhood. As usual in mean-field analyses, the focus is on qualitatively analyzing the evolution, characterizing the emerging phases, and comparing with simulation results to achieve a more complete understanding of the model. 

We find that our MFT is capable of maintaining and/or forming ordered states, such as bands and traffic jams, that coexist with a disordered background. Furthemore, it agrees qualitatively with simulations, showing similar behaviors for phase transitions and evolving into the same ordered pattern under the same conditions.

The remainder of this article is organized as follows. In Section~\ref{sec:model}, we introduce the three-state active lattice gas with excluded volume, describing in detail the dynamics of both the particle model and the MFT, as well as the analytical approach used to study condensed structures. Section~\ref{sec:result} presents the results obtained via MFT. In Section~\ref{sec:comp}, we compare our results with simulations, and in Section~\ref{sec:conclusion} we detail our conclusions.
%-------------------------------------------------------------------------------------------------%

\section{\label{sec:model} Model}
\begin{figure}[h!]
    \centering
    \includegraphics[width=0.65\linewidth]{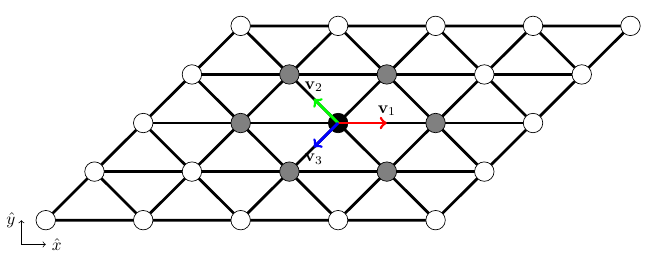}
    \caption{Schematic of the model. The neighborhood of the central site (black) corresponds to the site itself and the six first neighbors (gray). The velocities $\mathbf{v}_1$, $\mathbf{v}_2$, and $\mathbf{v}_3$ are depicted by the arrows in red, green, and blue, respectively.} 
    \label{fig:lattice}
\end{figure}

\textit{Particle model}. We consider a triangular lattice with periodic boundary conditions as shown in Figure~\ref{fig:lattice}. The distance between two neighboring sites is $a=1$, and the total number of sites is $L_x\mathrm{x}L_y$, which also defines the lattice size.

Each site can be occupied by at most one particle. Velocity directions are restricted to three orientations, as follows: 
\begin{equation}
    \label{eq:velocities}
    \begin{split}
        &\mathbf{v}_1 = v_0 \mathbf{\hat{x}}, \\ 
        &\mathbf{v}_2 = v_0\left(-\frac{1}{2} \mathbf{\hat{x}} + \frac{\sqrt{3}}{2} \mathbf{\hat{y}}\right),  \\ 
        &\mathbf{v}_3 = v_0 \left(-\frac{1}{2} \mathbf{\hat{x}} - \frac{\sqrt{3}}{2} \mathbf{\hat{y}} \right),
    \end{split}
\end{equation}
where $v_0$ denotes the speed of motion. A particle can access any site in the lattice via a sequence of steps along these three directions.

Let $N$ denote the number of particles; their states are defined by their position and  orientation. Let $N_i$ be the number of particles with velocity $\mathbf{v}_i$, such that $N = \sum_{i=1}^{3} N_i$. We define the fraction of particles in each state as $f_i = \dfrac{N_i}{N}, \ i=1,\ 2,\ 3,$, leading to the constraint $f_1+f_2+f_3 = 1.$

The orientation fractions reflect the degree of ordering. For example, $f_i > f_j,\ f_k$, signals orientational order, whereas $f_1=f_2=f_3=1/3$ corresponds to its absence. These fractions vary over time due to the alignment interaction between neighbors, in which a particle with some velocity $\mathbf{v}_i$ can reorient itself to align with the majority direction within its neighborhood. Here we consider a neighborhood formed by the first six neighbors of a particle, plus the particle itself, totaling seven sites involved in the alignment. By contrast, previous simulations of the active lattice gas \cite{rosembach2024} use an extended neighborhood, including up to third  neighbors for a total of nineteen sites. A smaller neighborhood was chosen to simplify the derivation of the equations of motion and the validation of numerical results in the mean-field approach, as well as to substantially reduce computational cost.

The configuration of the system is specified by the set of positions and orientations of the $N$ (conserved) particles. The evolution is sequential, involving, at each time step, the possible reorientation and movement of a randomly-chosen particle. A series of $N$ steps comprises one \textit{time unit}. At each step of the simulation algorithm, one of the $N$ particles is selected at random. The set of velocities among the occupied sites in the 7-site neighborhood centered on the selected particle is then examined. If a majority orientation exists, then with probability $1-\eta$ the central particle adopts this orientation, and with probability $\eta$ adopts, at random, one of the nonmajority orientations. If there is no majority (i.e., there is a two- or three-way tie), the central particle adopts, at random, one of the three possible orientations. After the alignment step, the particle attempts to move to the adjacent site in the direction of its updated velocity; if that site is empty, the particle moves there, while if its occupied, the particle remains in its current position, with its current updated orientation.  

\textit{Mean-field description}. The MFT description works with the set of occupation probabilities, $\rho_s$, and orientation fractions, $f_{i, s}$ of each site $s$. Thus, the probability of occupancy by a particle with orientation $i$ at site $s$ is $\rho_s f_{i,s}$. Each site is in one of four possible states: empty with probability $p_0 = (1-\rho_s)$; occupied with velocity $\mathbf{v}_1$, $\mathbf{v}_2$, or $\mathbf{v}_3$ with probabilities $p_1 = \rho_s f_{1,s}$, $p_2 = \rho_s f_{2,s}$, and $p_3 = \rho_s f_{3,s}$, respectively. Accordingly, the mean-field approximantion considers a set of three independent density fields, $p_1$, $p_2$, and $p_3$, corresponding to the three possible orientations; $p_0$ is not considered a density field since its value is derived from $\rho_s$. The average of the local densities provides the global lattice density

\begin{equation}
    \label{eq:total_density}
    \rho_g = \frac{1}{A}\sum_{s=1}^{A} \rho_s,
\end{equation}
with $A=L_x\mathrm{x}L_y$.

\begin{figure}[h!]
    \centering
    \includegraphics[width=0.4\linewidth]{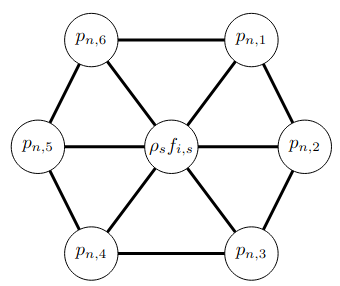}
    \caption{The neighborhood of a particle with associated occupancy probabilities. The central site is occupied and in state $i$, while its neighbors may be empty or occupied. The occupancy probabilities $p_n$ for the sites in the neighborhood are assumed to be mutually independent.}
    \label{fig:neighborhood}
\end{figure}

The dynamics consists of alternating alignment and dislocation steps. The alignment step is modeled by equations of motion for the fractions of particles at each state, accounting for reorientation toward the majority direction and disturbance by noise. To formulate these equations, we examine the neighborhood of a site occupied by a particle in state $i$ and consider the four possible events that can occur there:

\begin{itemize}
    \item $M=0$ - There is no majority;
    \item $M=j$ - The majority of particles have velocity $\mathbf{v}_j$, with $j =1, 2, 3$.
\end{itemize}
Let $\mathcal{P}\left(M,i\right)$ denote the probability that event $M$ occur in the 7-site neighborhood centered on site $s$, given that it is in state $i$. Event $M$ corresponds to various configurations in this neighborhood, each with its own probability $\mathcal{P_C}$. To calculate this probability, we assume, as usual in mean-field analyses, that the occupancy probabilities of each site in the neighborhood are mutually independent; thus, $\mathcal{P_C}$ can be factorized and expressed as:
\begin{equation}
    \label{eq:prob_conf}
    \mathcal{P_C} = \rho_s f_{i, s} \prod_{k=1}^{6} p_{n,k},
\end{equation}
where $p_n$, with $n=0,\ 1,\ 2,\ 3$, represents the occupancy probability of site $k$ in the neighborhood (see Figure~\ref{fig:neighborhood}). 
The probability of event $M$ is then obtained by summing over all configurations that map onto it,
\begin{equation}
    \label{eq:prob_tot}
    \mathcal{P}\left(M, i\right) = \sum_{\mathcal{C} \subset (M,i)} \mathcal{P_C}.
\end{equation}

The probabilities in Eq.~(\ref{eq:prob_tot}) would determine the likelihood of reorientation of the central particle if there were no noise. The noise $\eta$ represents the probability that a particle will perceive the majority incorrectly. Consider a particle in state $i$ in a neighborhood where event $M$ occurs. The effect of the noise follows these rules:
\begin{itemize}
    \item If $i=M > 0$: The probability $f_{i,s}$ is reduced by $\eta$, with an equal fraction $\eta/2$ transferred to the probabilities of the minority states $f_{j,s}$ and $f_{k,s}$;   
    \item If $i \ne M$: A probability $(1-\eta)$ is transferred from $f_{i,s}$ to the probability of the state corresponding to event $M$ at site $s$;  and a probability of $\eta/2$ is redistributed to the alternative minority state;
    \item If $M=0$: The probability $f_{i,s}$ is reduced by $2\eta/3$, with an equal fraction $\eta/3$ transferred to the probability of the other two states $f_{j,s}$ and $f_{k,s}$.
\end{itemize}
For example, a particle in state $i$ with exactly two occupied neighbors, in states $j \neq k \neq i$ ($M=0$) updates its orientation randomly, whereas an isolated particle in state $i$ ($M=i$) maintains its orientation with probability $1-\eta$ and adopts state $j \neq i$ or $k \neq i$ each with probability $\eta/2$.

Given $\mathcal{P}\left(M, i\right)$ and the noise rules, the equations of motion for the fractions of particles in each state are:
\begin{widetext}
\begin{equation}
    \label{eq:eq_motion}
    \begin{split}
    \frac{df_i}{dt} = -&f_i \left\{\eta \ \mathcal{P}\left(i, i\right) + \left(1- \frac{\eta}{2} \right) \left[\mathcal{P}\left(j, i\right) + \mathcal{P}\left(k, i\right) \right] + \frac{2}{3}\eta \ \mathcal{P}\left(0, i\right)\right\} \\
    + &f_j \left \{\left(1-\eta\right) \mathcal{P}\left(i, j\right)+ \frac{\eta}{2} \left[\mathcal{P}\left(j, j\right) + \mathcal{P}\left(k, j\right)\right] + \frac{\eta}{3} \mathcal{P}\left(0, j\right) \right\} \\
    + &f_k \left \{\left(1-\eta\right) \mathcal{P}\left(i, k\right)+ \frac{\eta}{2} \left[\mathcal{P}\left(j, k\right) + \mathcal{P}\left(k, k\right)\right] + \frac{\eta}{3} \mathcal{P}\left(0, k\right) \right\},
    \end{split}
\end{equation}
\end{widetext}
where $i$, $j$, and $k$ take on the values $1$, $2$, and $3$, in cyclic order. Each site has its own equations of motion, which provides the variations of the fractions due to the alignment interaction. These equations are highly nonlinear and therefore require numerical integration; for this, we use the fourth-order Runge-Kutta scheme to obtain the values of $f_i$ at each time step $\Delta t$. 

Specifying the noise intensity $\eta$ and the occupancy probability $p_n$ for each possible event at the initial time provides a sufficient set of boundary conditions to determine the time evolution of $f_i$ at each site in the lattice. The closed set of equations for the density fields actually consists of only two equations of motion due to the normalization constraint on the fractions, since the third component is uniquely determined by $\sum_i f_i = 1$. In the present study, however, we opted for simultaneous numerical integration of the three differential equations to ensure that the identity is preserved over time.

Once the values of $f_i$ have been updated, particles can move through the lattice via a density transfer process with excluded volume. The density, $\Delta g = \Delta\rho_s f_{i,s}$, transferred from $\mathbf{s}$ to the neighbor $\mathbf{s'} = \mathbf{s} + \mathbf{v}_i$ during time interval $\Delta t$ depends on the current density, $g_0$, and on the density, $u_0$, its neighbor can receive without violating the constraint, $0 \le \rho_s \le 1$. We have,
\begin{equation}
    \label{eq:density_transfer}
    \begin{split}
        \Delta g = \frac{v_0 \Delta t g_0^2}{v_0\Delta t g_0 - 1},\qquad \qquad \quad &\mathrm{if}\ g_0=u_0, \\
        \Delta g = - \frac{g_0\left[e^{v_0\Delta t(g_0 - u_0)} - 1\right]}{e^{v_o \Delta t (g_0-u_0)} - g_0/u_0}, \quad &\mathrm{if}\ g_0 \ne u_0.
    \end{split}
\end{equation}
A derivation of Eq.~(\ref{eq:density_transfer}) is provided in Appendix A. Notice that we have included the particle speed $v_0$ in the definition of the density transfer, which is restricted to nearest neighbors regardless of the value of $v_0$.

The spatial structure of the lattice is incorporated into the MFT through Eq.~\ref{eq:density_transfer}, which governs the occupation variation of a site based on the density flows received from and sent to its neighbors at each instant $\Delta t$. Note that, due to the geometric limitations of velocity orientations, a site does not receive particles from the same neighbors to which it can send particles. The densities updated by Eq.~\ref{eq:density_transfer} supply the occupation probabilities that constitute Eq.~\ref{eq:eq_motion}; this, in turn, considers the occupation of all sites in the neighborhood to model the alignment interaction. Furthermore, in the equations of motion, all possible configurational combinations in a given neighborhood are computed. For example, the probability that a particle in state $i$ belongs to a neighborhood where event $i$ occurs ($\mathcal{P}\left(i, i\right)$) must account for the scenarios in which all neighbors are empty, expressed by $\prod_s (1-\rho_s)$, and the opposite limit in which all neighboring sites are occupied, given by $\prod_s \rho_s$ (considering the various states that lead to event $i$), as well as all intermediate configurations. Thus, the local density of sites enters directly in the equations of motion, altering the statistical weight of specific configurations. Consequently, $\rho_s \approx 1$ implies that site $s$ is more likely to induce the alignment of its neighbors, while also preventing particles from moving toward it, thereby promoting order in dense regions; in contrast, $\rho_s \approx 0$ indicates a weakened influence on local reorientation, which contributes to maintaining a disordered phase in low-density areas.

While simulations are performed using random sequential updating, the MFT framework employs parallel updating, where all particles undergo the alignment process followed by all possible displacement processes. The model parameters are the lattice size $L_x\mathrm{x}L_y$, velocity $v_0$, global density $\rho_g$, and noise intensity $\eta$; the initial condition is defined by the values given for the fractions of particles, $f_{i,s}$, and densities, $\rho_s$, for all sites at time $t=0$. At each time step $f_{i,s}$ and $\rho_s$ are updated for the entire lattice. This cycle of alignment and movement repeats, updating neighborhoods, until a steady state is reached, defined by no further variation in the average orientation fractions.

An ordered phase is characterized by the predominance of one orientation over a region, or the entire system. Conversely, the disordered phase corresponds to the case in which the density is homogenous and all three fractions are equal, $f_i = 1/3,\ \forall\ i$.

We define the orientational order parameter as
\begin{equation}
    \label{eq:ord_par}
    \Phi = \sqrt{f_1^2 + f_2^2 + f_3^2 - f_1 f_2 - f_1 f_3 - f_2 f_3},
\end{equation}
where the $f_i$ are the fractions averaged over the lattice. This definition resembles the nematic order parameter \cite{plishke}. When $\Phi = 0$, the system is in an isotropic disordered phase, whereas for $\Phi > 0$ a dominant direction emerges, indicating orientational order.

%-------------------------------------------------------------------------------------------------%

\section{\label{sec:result} Results}

The results of the mean-field theory presented here focus mostly on the condensed structures reported in \cite{rosembach2024}. An \textit{immobile band} (IB) is characterized by a dense region that spans the lattice and exhibits a well defined orientation. The preferred orientation of the particles is parallel to the band, which is a low-mobility structure due to the excluded volume preventing the flow of particles inside the band. On the other hand, a \textit{mobile band} (MB) refers to a dense and ordered region that does not span the lattice, so that, in principle, it can move through the system in the direction of the majority velocity, while in the bulk of the structure the movement is inhibited by high occupation.

\textit{Traffic jams} are congested regions of particles moving in different directions, leading to mutual blockage. This is a typical pattern in models with excluded-volume interactions \cite{rosembach2024, peruani2011, karmakar2022}. In this work, we distinguish two types: \textit{type-I traffic jams} (TJ-I), in which clusters of particles with the three possible orientations encounter each other; and \textit{type-II traffic jams}  (TJ-II), in which the low mobility is due to the meeting of only two opposing directions. Note that a TJ-I is a compact, localized structure, while a TJ-II spans the lattice.

\begin{figure}[h]
    \centering
    \includegraphics[width=0.55\linewidth]{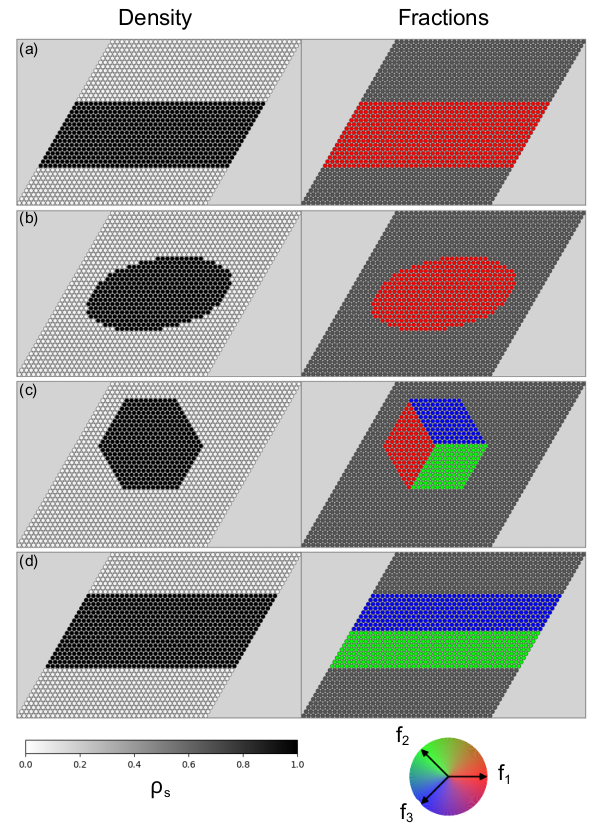}
    \caption{\label{fig:InitialConditions}Initial configurations representing condensed structures. (a) Immobile band with $\mathbf{v}_1$ as majority. (b) Mobile Band with $\mathbf{v}_1$ as majority. (c) type-I Traffic Jam. (d) type-II Traffic Jam with $\mathbf{v}_2$ and $\mathbf{v}_3$ dominating. The density plots have a gray map color-scale; $f_1$, $f_2$, and $f_3$, are represented by red, green, and blue, respectively, with disordered regions rendered in gray. In the numerical integration we use $v_0 = 0.5$ in cases (a) and (d), and $v_0 = 0.1$ in cases (b) and (c).}
\end{figure}

To examine the robustness of a given condensed structure, the lattice is initialized with configurations that replicate its characteristic features. We define a high-density ordered region, consisting of particles with aligned orientations for the low-mobility area. The remaining sites are initialized in a disordered state, with random orientations and lower densities, representing the coexistence of ordered and disordered regions. Figure~\ref{fig:InitialConditions} illustrates the initial configurations (ICs) constructed to study the immobile band, mobile band and traffic jam patterns. For more details on these structures see Appendix B.

We employ a lattice of size $L_x =  L_y = 40$, and speed set to $v_0=0.1$ or $v_0 =0.5$. The difference in the value of $v_0$ is due to the fact that, for some ICs the higher value of $v_0 = 0.5$ generated non-physical results. For each value of global density $\rho_g$, defined by the initial distribution of the site densities, we performed calculations over a range of noise values. Time evolution is computed using the fourth-order Runge-Kutta method with a time step of $\Delta t = 0.01$; integrations are carried out over one million iterations to ensure that the system reaches a stationary state. Below we report the kinds of final configurations observed as we vary density and noise intensity; each subsection treats a different kind of initial configuration.

\subsection{\label{subsec:IB}Immobile Band}

In order to study the necessary conditions for the IB pattern to form and/or persist during the evolution, we analyzed three initial configurations:
\begin{itemize}
    \item [I -] A dense, horizontally oriented band with all particles with velocity $\mathbf{v}_1$, as shown in Figure~\ref{fig:InitialConditions}(a);
    \item[II -] Homogeneous density across the lattice, with a localized region where all particles move with velocity $\mathbf{v}_1$ forming a horizontal band - outside this region, the system remains in a disordered state;
    \item[III -] A dense horizontal band with $f_1 = 1/3 + \delta$, and $f_2=f_3=1/3-\delta/2$, with $\delta = 10^{-3}$.
\end{itemize}

The study using IC I examined the stability of a fully-formed IB, while IC III tests the susceptibility of a dense band to a small asymmetry in orientational populations. In IC II, the ability of an IB to spread into a dense exterior is evaluated. The evolution is analyzed for different noise values at four global densities. At lower densities, $\rho_g = 0.23$ and $\rho_g = 0.31$, the initial bands are smaller in comparison with the higher-density cases, $\rho_g = 0.43$ and $\rho_g = 0.52$. In addition to the width of the bands, the lattice density can be varied by changing the probability of occupation of the sites, $\rho_s$, in the disordered region, but keeping in mind that this must be a region with low occupancy.

\begin{figure}[!htp]
    \centering
    \includegraphics[width=0.65\linewidth]{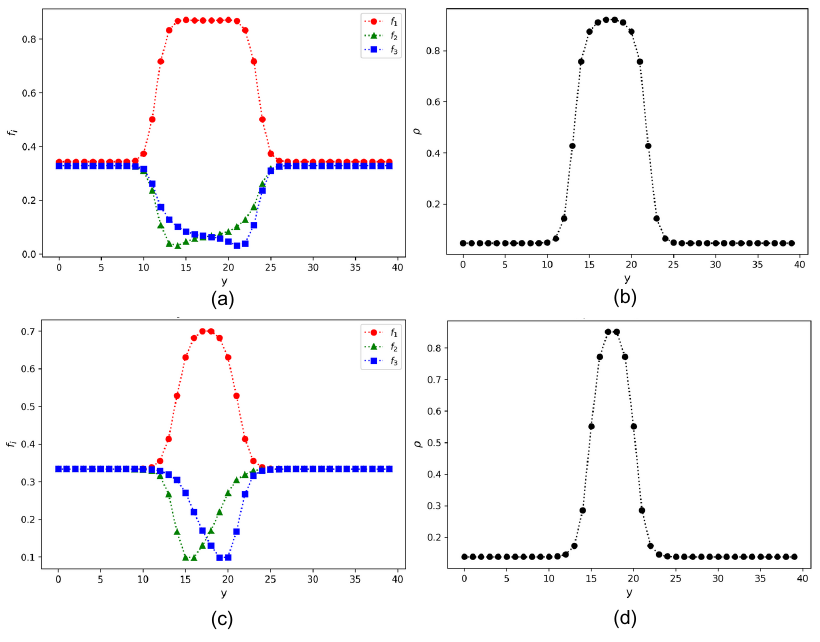}
    \caption{\label{fig:perfil_IB023} Profiles of orientation fractions $f_i$ and densities $\rho_s$ along the $y$ direction for the IB in the steady state for global density $\rho_g=0.23$. Panels (a) and (b) show the profiles for noise intensity $\eta = 0.01$; (c) and (d) are for $\eta = 0.045$, near the transition to the disordered state.}
\end{figure}

For $\rho_g = 0.23$, the IB with $\mathbf{v}_1$ as a majority forms quickly (for ICs II and III) and persists (for all ICs) for $\eta < \eta_c$. The IB structure only disappears with the transition to the disordered phase in which the system is homogeneous. In Figure~\ref{fig:perfil_IB023} we show the fractions and density profiles along the $y$ direction for the steady state for two noise values with IC-I; for $\eta=0.01$ the final dense band is similar to the one in the initial condition in size and densities, however as the noise increases the band becomes smaller and less populated, as shown for $\eta = 0.045$.

 The behavior for $\rho_g = 0.31$ with ICs I and II is similar to that found for $\rho_g = 0.23$. However, IC III, with $f_1$ slightly larger than the other fractions, does not lead to formation of an IB structure. Instead, we observe the emergence of a type-II traffic jam with $\mathbf{v}_2$ and $\mathbf{v}_3$ as the majority within the dense band, as can be seen in the Supplementary Material \cite{supp}.

\begin{figure}[!htp]
    \centering
    \includegraphics[width=0.65\linewidth]{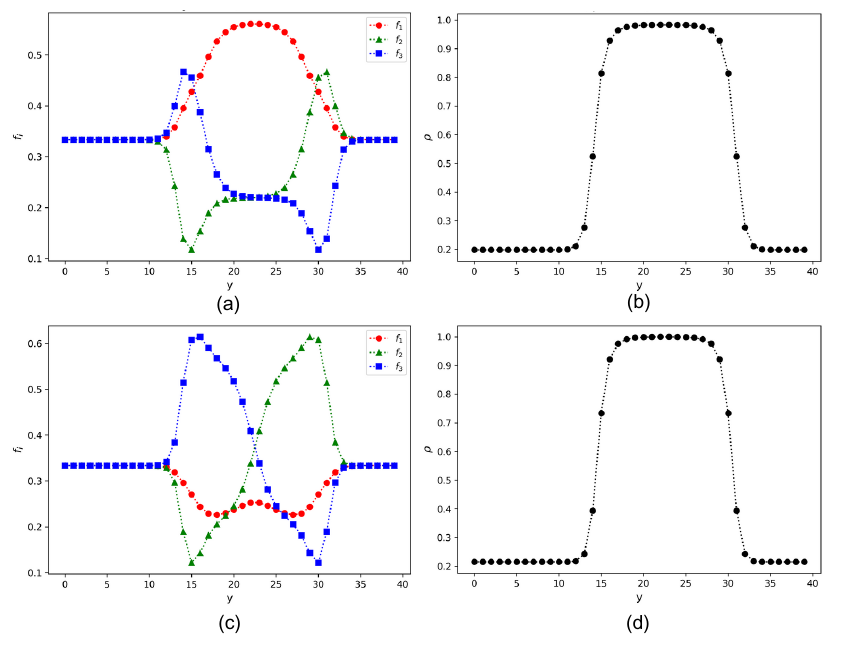}
    \caption{\label{fig:perfil_IB052} Profiles of orientation fractions $f_i$ and densities $\rho_s$ along the $y$ direction for the IB and TJ-II in the steady state for global density $\rho_g=0.52$. Panels (a) and (b) show the profiles for $\eta = 0.096$, when the final configuration is still an IB but near the transition to TJ-II; (c) and (d) are for $\eta = 0.1$, for which the final configuration is a TJ-II.}
\end{figure}

\begin{figure}[!htp]
	\center
	\includegraphics[width=0.5\linewidth]{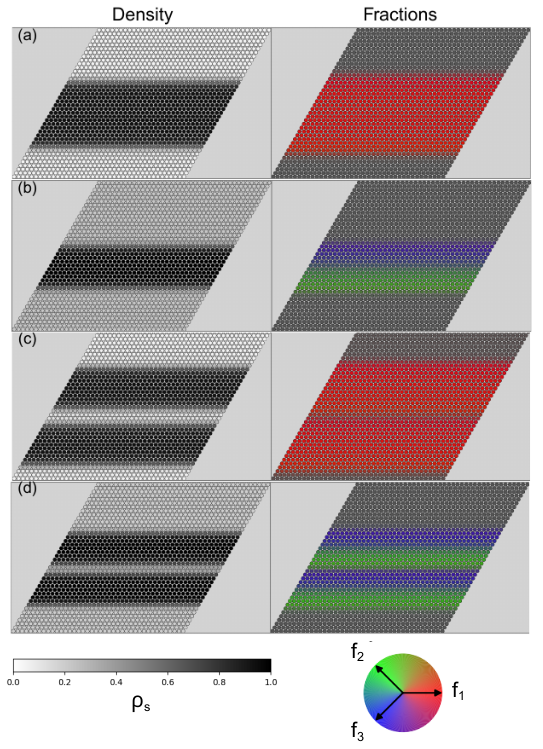}
	\caption{\label{fig:IB_Finals} Final configurations for the Immobile Band studies: (a) IB with $\mathbf{v}_1$ as majority for $\rho_g = 0.43$, $\eta=0.01$ and IC I. (b) TJ-II with $\mathbf{v}_2$ and $\mathbf{v}_3$ dominating for $\rho_g = 0.43$, $\eta = 0.1$ and IC I. (c) Emergence of two IBs for $\rho_g=0.52$, $\eta = 0.01$ and IC II. (d) Emergence of two TJ-IIs for $\rho_g=0.52$, $\eta = 0.08$, and IC II.}
\end{figure}

More diverse results appear for higher densities. For ICs I and II, the IB is maintained for small $\eta$, as seen in Fig.~\ref{fig:IB_Finals}(a). However, as the noise increases, there is a transition to a TJ-II, with the orientation $\mathbf{v}_1$ losing the majority within the band. Figure~\ref{fig:perfil_IB052} shows the profiles for $\rho_g = 0.52$ near the transition for $\eta = 0.096$, when the steady state still exhibits an IB with $\mathbf{v}_1$ as majority, but with particles with velocity $\mathbf{v}_2$ and $\mathbf{v}_3$ crossing the borders of the band. This picture also shows the profile for $\eta=0.1$, for which the stationary state is a TJ-II. Figure~\ref{fig:IB_Finals}(b) shows that in the new steady state, the dense band is equally divided into two smaller bands, the upper with majority $\mathbf{v}_3$, and the lower with majority $\mathbf{v}_2$, as expected since, in this arrangement, the two velocities are mutually blocking. For IC II we also see the emergence of two regions with IB or TJ-II for some noise values \cite{supp}, as shown in Figures \ref{fig:IB_Finals}(c)-(d); this suggests one could have even more bands with IB or TJ-II, if the system were larger and/or denser. For IC III we observe the same behavior as described for $\rho_g=0.31$: the structure formed is a TJ-II for any noise value below the transition to the disordered phase. 

The results for IC III persist even for a disordered initial state, i.e., for $\delta = 0.0$. This suggests that, under these conditions, the emergence of a condensed structure is driven by density variations rather than by fluctuations in the particle fractions. These findings, together with the analysis presented in the Supplementary Material regarding density fluctuations \cite{supp}, indicate that the system tends to develop a low-mobility structure in the high-density region. In this case, the type-II traffic jam appears as the preferred pattern. However, for $\rho_g = 0.23$, the initial dense band is too narrow to undergo the partitioning required for the formation of a TJ-II, leading instead to the emergence of an IB with velocity $\mathbf{v}_1$.

\begin{figure}[!htp]
    \centering
    \includegraphics[width=0.6\linewidth]{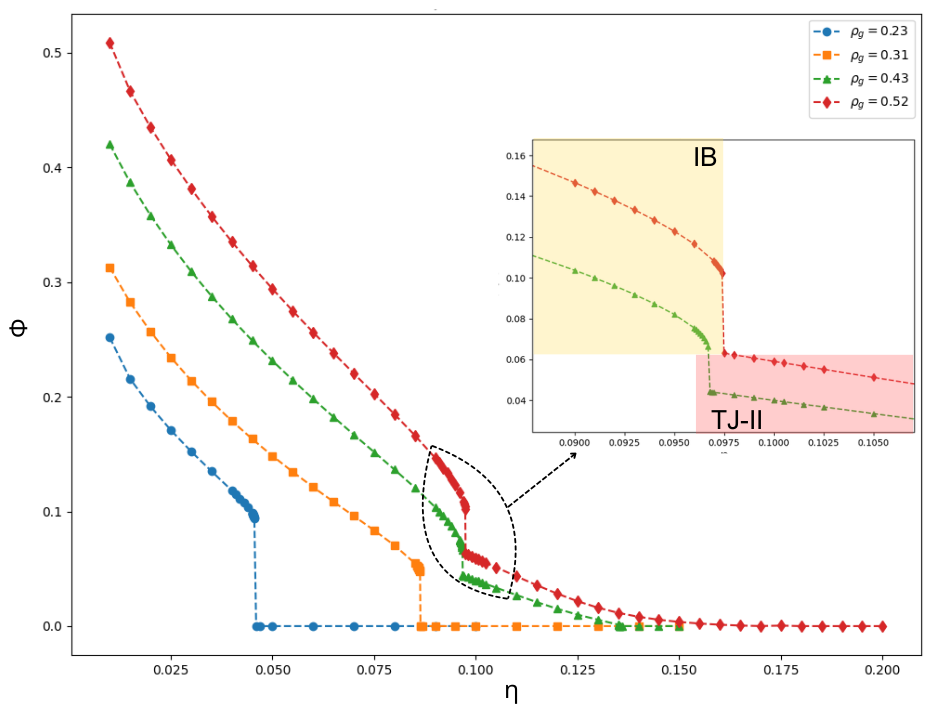}
    \caption{\label{fig:ord_par_densities} Order parameter as a function of noise for IC I. The inset highlights the transition to TJ-II that occurs for $\rho_g = 0.43$ and $\rho_g = 0.52$: the yellow-shaded area indicates the region with IB as final configuration, while the red-shaded area indicates that TJ-II appears in the final configurations. TJ-II has a smaller value of $\Phi$ than IB due to the shared majority between the two velocities.}
\end{figure}

Figure~\ref{fig:ord_par_densities} shows the phase transitions that occur for IC I: for low densities, in which the transition to TJ-II does not occur, the transition to the disordered state in our mean-field theory is discontinuous; for high densities we have a discontinuous transition to TJ-II, followed by a continuous transition to the disordered phase.
We also see that the higher the density, the greater the noise intensity required to disorder the system. 

\begin{figure}[!htp]
	\center
	\includegraphics[width=0.8\linewidth]{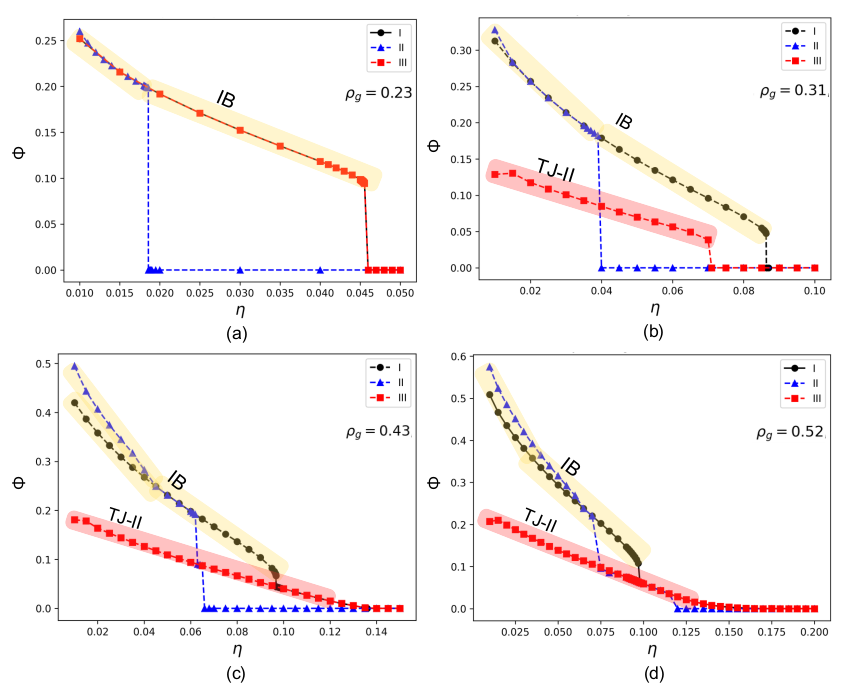}
	\caption{\label{fig:IB_ordpar} Order parameter as a function of noise for Immobile Band ICs. Each point is computed independently by starting from the initial condition at the specific $\eta$ value shown for different densities: (a) $\rho_g=0.23$, where the black circle representing IC I is hidden behind the red squares for IC III, indicating that both configurations lead to the same final state. (b) $\rho_g = 0.31$. (c) $\rho_g = 0.43$. (d) $\rho_g = 0.52$. The yellow-shaded areas indicate the conditions in which the steady state is an IB, while the pink-shaded ones indicate the conditions for which the steady state is a TJ-II. The lower values of $ \Phi $ for IC III (red square) in the last three figures correspond to the formation of a TJ-II instead of an IB. For higher densities, the transition to TJ-II that occurs for ICs I and II is marked by a jump to smaller values of $ \Phi $. }
\end{figure}

From Figure~\ref{fig:IB_ordpar} we see that, at $\rho_g = 0.23$, the ICs with a dense horizontal band are more stable than IC II with uniform density. The same applies when comparing ICs I and II at the higher densities, with the distinction that IC III now evolves to a TJ-II. For ICs II and III, the transitions to the disordered phase continue to be discontinuous, including $\rho_g=0.31$ with IC III where no IB is formed (see Fig.~\ref{fig:IB_ordpar}(a)-(b)). Figures~\ref{fig:IB_ordpar}(c)-(d) further show that, for IC II at high densities, the transitions to TJ-II and subsequently to the disordered state are discontinuous, whereas IC III exhibits a continuous transition to the disordered state.

\newpage
\subsection{\label{subsec:MB}Mobile Band}

The MB initial condition is built as a dense oval cluster where all particles have velocity $\mathbf{v}_1$, see Figure~\ref{fig:InitialConditions}(b). Whilst mobility is low within the band, the structure itself is free to move through the low-density disordered background, as particles escape from the leading edge into the low-density region, and accumulate at the trailing, high-denisty edge. By varying the density of the disordered region we were able to analyze three values of global densities, $\rho_g=0.29,\ 0.33,\ \mathrm{and}\ 0.37$.

\begin{figure}[!htp]
	\center
	\includegraphics[width=0.65\linewidth]{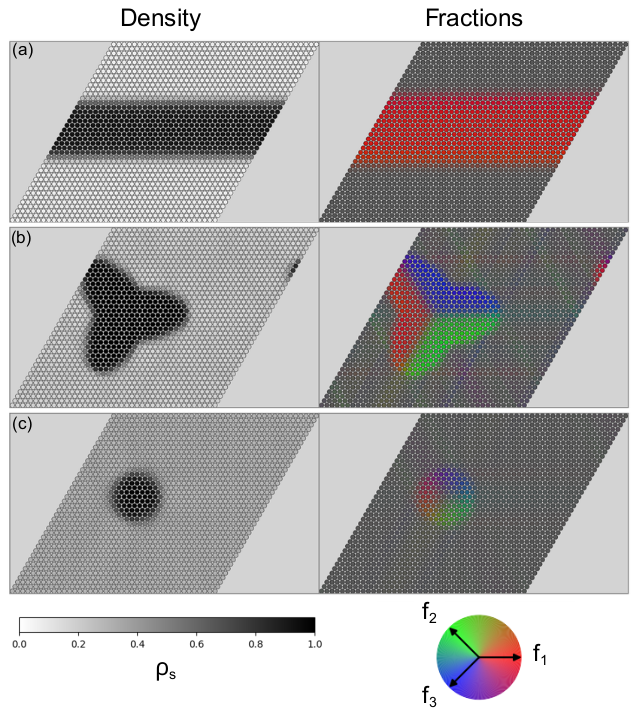}
	\caption{\label{fig:MB_Finals} Final configurations for the Mobile Band studies with $\rho_g = 0.33$: (a) IB with $\mathbf{v}_1$ as majority for $\eta=0.02$. (b) TJ-I for $\eta = 0.06$. (c) TJ-I for $\eta = 0.12$.}
\end{figure}

The behavior is as follows: for small noise values, the initial mobile band expands across the lattice, forming an immobile band, as shown in Figure~\ref{fig:MB_Finals}(a); for $\eta \simeq 0.06$ for the three densities, the system goes through a transition in which a type-I traffic jam forms in a shape resembling a three-leaf clover, as depicted in Figure~\ref{fig:MB_Finals}(b) \cite{supp}. As $\eta$ increases the TJ-I structure shrinks (see Fig.~\ref{fig:MB_Finals}(c)), and eventually the transtition to the disordered state occurs. 

The TJ-I pattern that emerges in our model has the distinctive feature of appearing disordered at the lattice scale. This occurs because the dense region is evenly divided into three domains, each dominated by a distinct velocity orientation, and surrounded by a disordered background. Consequently, we obtain $\langle f_i \rangle = 1/3,\ \forall i$, which implies that the order parameter vanishes, $ \Phi  = 0$, even though the system is in an organized state. To gain a deeper understanding of the behavior and the transition for this configuration, we introduce a new parameter:
\begin{equation}
    \label{eq:deviation}
    \tilde{\sigma} = \frac{1}{\rho_g A} \sum_{i=1}^{L_x} \sum_{j=1}^{L_y} \rho^{ij}\sqrt{\sum_{k=1}^{3} \mid f_{k}^{ij} - \frac{1}{3} \mid ^2}\ .
\end{equation}
$\tilde{\sigma}$ quantifies the deviation from the orientationally disordered state, normalized by the system size and weighted by the local density. This quantity reveals the presence of well-aligned regions in a system where the global order parameter vanishes.

\begin{figure}[!htp]
    \centering
    \includegraphics[width=0.65\linewidth]{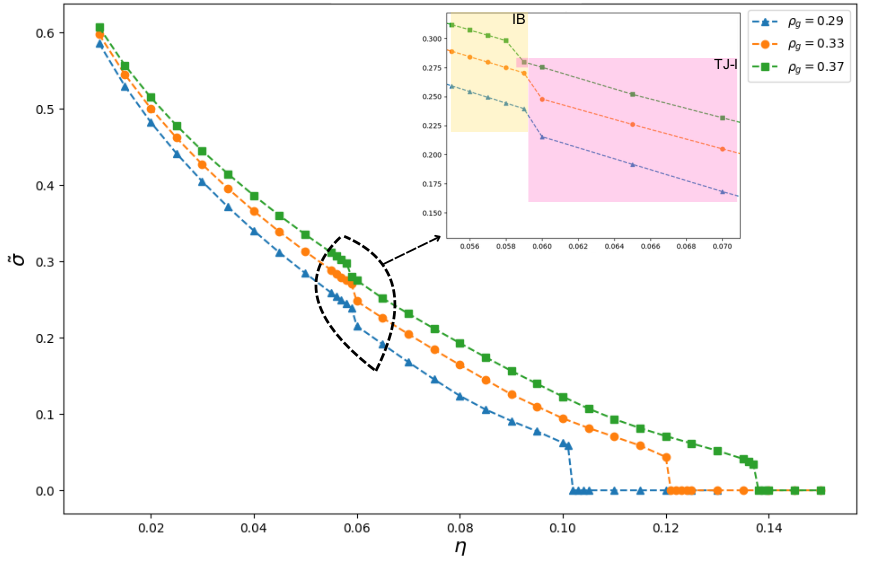}
    \caption{\label{fig:deviation_MB} Deviation of the fractions from the orientationally disordered state as a function of noise for Mobile Band initial conditions. The inset highlights the transitions from IB to TJ-I that occur for all three densities: the yellow-shaded area indicates the region where IB emerges as the final configuration, while the pink-shaded area indicates that TJ-I appears in the final configurations. The discontinuous transition to the disordered state occurs when $\tilde{\sigma}$ vanishes.}
\end{figure}

Figure~\ref{fig:deviation_MB} shows the deviation $\tilde{\sigma}$ as a function of noise for the three densities analyzed. The value of $\tilde{\sigma}$ is higher the greater the degree of order within the lattice, so as $\eta$ increases, the deviation of the fractions from the disordered state decreases. The transition from IB to TJ-I is characterized by a small drop in the value of $\tilde{\sigma}$ at $\eta \simeq 0.06$; after the transition, the deviation continues to decrease, indicating the shrinkage of the ordered area in the lattice. Finally, the disordered state is reached when $\tilde{\sigma} = 0$, that is, $f_i = 1/3,\ \forall i$, at all sites. 

\subsection{\label{subsec:TJIsym}Type-I Traffic Jam: Symmetric Structure}

The initial configuration constructed to analyze the type-I traffic jam pattern, as depicted in Figure~\ref{fig:InitialConditions}(c), is equally divided into regions with majorities $\mathbf{v}_1$, $\mathbf{v}_2$, and $\mathbf{v}_3$. The pattern is built in such a way that the orientations of the velocities block each other, guaranteeing low mobility along with high density. While the dense region is fully occupied, the disordered one assumes different values for the density, allowing us to vary the global density. Thus, we studied the behavior for $\rho_g = 0.27,\ 0.30,\ \mathrm{and}\ 0.34$.

For all the global densities studied, allowing the system to evolve from the IC described above, we observe that the structure of a TJ-I persists, while the initial shape changes to that of a three-leaf clover \cite{supp}, as seen in Sec.~\ref{subsec:MB} with the emergence of a traffic jam from a mobile band. For low noise, the leaves of the traffic jam stretch across the lattice so that the ordered region becomes larger than in the initial configuration, as shown in Figure~\ref{fig:TJI_Finals}(a) - for very low noise, a rotation of the ordered structure occurs (see the Supplementary Material). The area of the ordered pattern becomes smaller as the noise increases (see Fig.~\ref{fig:TJI_Finals}(b)) until it completely disappears when $\eta_c$ is reached. From that point on, the steady state is always disordered.

\begin{figure}[!htp]
	\center
	\includegraphics[width=0.7\linewidth]{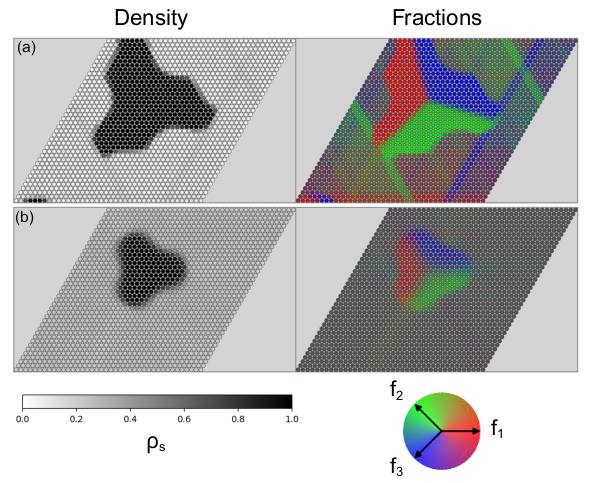}
	\caption{\label{fig:TJI_Finals} Final configurations for the type-I Traffic Jam studies for $\rho_g = 0.30$ with (a) $\eta=0.01$, and (b) $\eta = 0.08$. Despite apparent variations due to the periodic boundary conditions, both configurations are invariant under permutations of $\{1, 2, 3\}$ accompanied by an appropriate rotation of $2\pi/3$ about the center of the TJ-I.}
\end{figure}

\begin{figure}[!htp]
    \centering
    \includegraphics[width=0.6\linewidth]{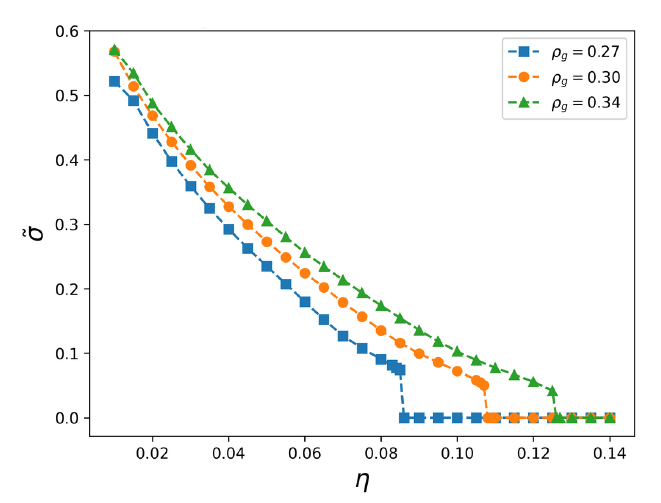}
    \caption{\label{fig:deviation_TJI} Deviation of the fractions from the orientationally disordered state as a function of noise for the symmetric type-I Traffic Jam as initial condition. The TJ-I persists as the steady state, until the transition to the disordered phase at $\eta_c$.}
\end{figure}

Figure~\ref{fig:deviation_TJI} shows the deviation from the orientationally disordered state $\tilde{\sigma}$ for different noise values. The decrease in the deviation parameter with increasing noise reflects the contraction of the traffic jam structure, which results in more sites in the disordered state. The only phase transition is of the order-disorder type, characterized by a sudden jump to $\tilde{\sigma} = 0$. We see that the higher the density, the greater the noise required to disorder the lattice. 

\subsection{\label{subsec:TJIasym}Type-I Traffic Jam: Asymmetric Structure}

According to our definition, the type-I traffic jam occurs when the three velocities are present within a dense region arranged in such a way that they mutually block one another. In this sense, there are different ways to implement this pattern on the lattice. We therefore study an asymmetric structure that fits the definition of a traffic jam but yields results significantly different from those discussed in the previous subsection.

\begin{figure}[!htp]
    \centering
    \includegraphics[width=0.85\linewidth]{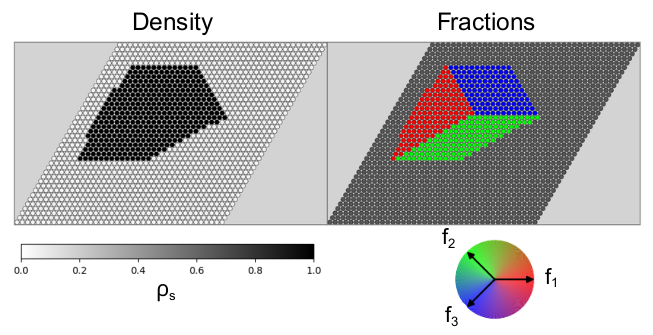}
    \caption{\label{fig:TJI-asy} Initial configuration representing the asymmetric type-I Traffic Jam.}
\end{figure}

The lattice is initialized with the configuration depicted in Figure~\ref{fig:TJI-asy}. As usual, the dense region is fully occupied and with a defined majority in each region, while the disordered region can assume different values of the density. Thus, we studied the behavior of this structure for $\rho_g = 0.25,\ 0.30,\ 0.32,\ 0.35,\ \mathrm{and}\ 0.40$.

\begin{figure}[!htp]
    \centering
    \includegraphics[width=0.7\linewidth]{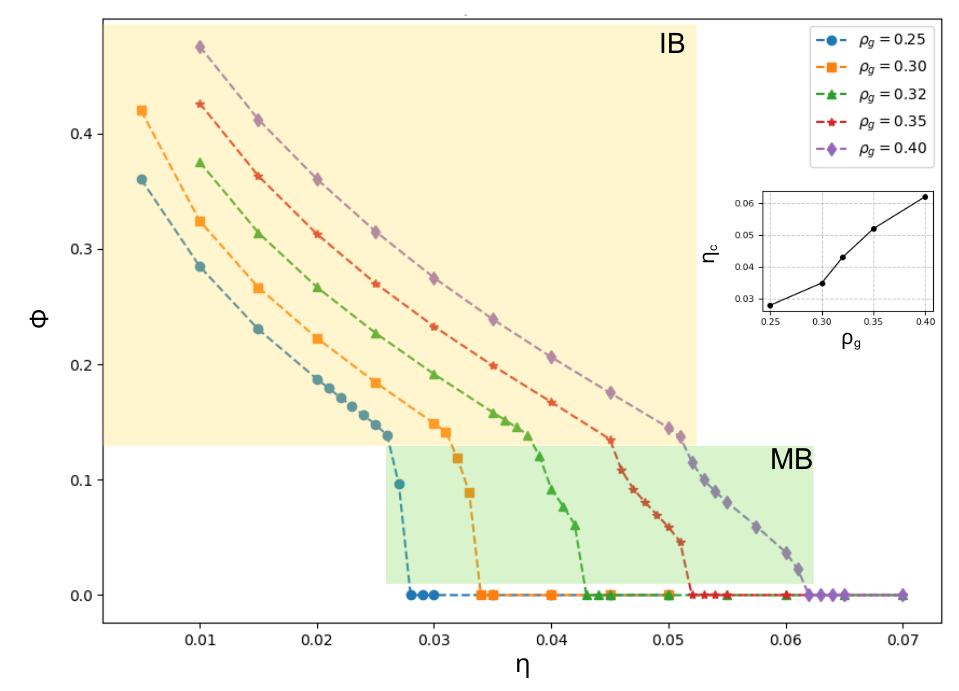}
    \caption{\label{fig:ordpar_TJIasym}Order parameter as a function of noise for the asymmetric TJ-I. The yellow-shaded area indicates the region where the final state is an IB, and the green-shaded one indicates the transition to MB. The inset shows the interplay between density $\rho_g$ and critical noise $\eta_c$: as $\rho_g$ increases, a higher noise is necessary to disorder the lattice.}
\end{figure}

The results are similar for all densities: the asymmetric TJ-I structure does not persist for any condition. For small noise, we see the formation of an IB, and, as the noise is increased, the band takes longer and longer to span the lattice and eventually we have the emergence of a MB in the steady state. The mobile band propagates through the lattice in a soliton-like manner \footnote{We have not studied collisions between MBs, but this remains an interesting question for future work.} - a cluster of particles moving periodically in the same direction, preserving its shape and maintaining a constant average particle fraction despite local variations due to displacement \cite{supp}. Thus, we obtain what is technically a nonsteady (periodic) solution to the equations for the site probabilities, even though the system reaches a globally stationary state. Figure~\ref{fig:ordpar_TJIasym} shows that the transition from IB to MB is characterized by a small drop in the value of the order parameter. For higher values of noise, there is a discontinuous transition to the disordered phase.

\begin{figure}[!htp]
	\center
	\includegraphics[width=0.55\linewidth]{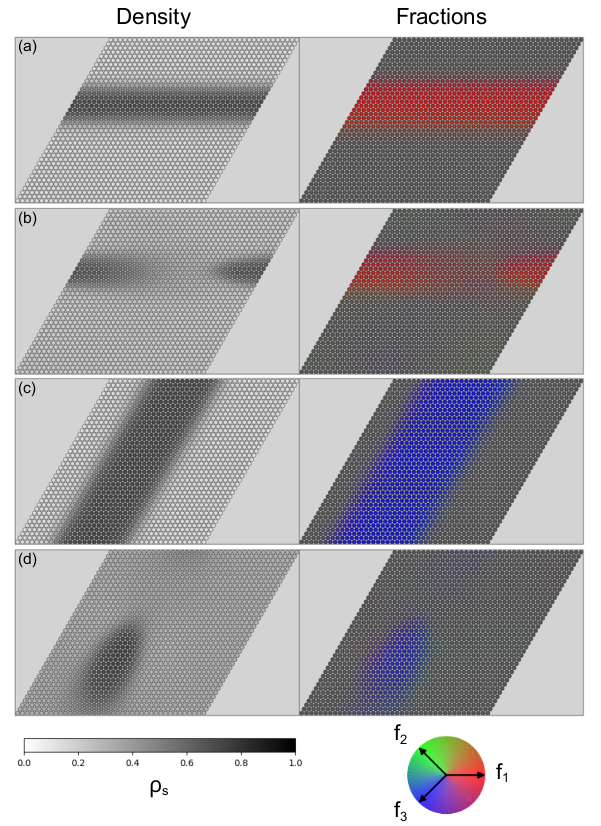}
	\caption{\label{fig:TJIasy_Finals} Final configurations for the asymmetric type-I traffic jam studies: (a) An IB with $\mathbf{v}_1$ as majority emerges for $\rho_g = 0.25$ and $\eta = 0.02$. (b) The MB in the direction of $\mathbf{v}_1$ for $\rho_g = 0.25$ and $\eta = 0.027$. (c) An IB with $\mathbf{v}_3$ as majority for $\rho_g = 0.35$ and $\eta = 0.02$. (d) A band moving in the direction of $\mathbf{v}_3$ for $\rho_g = 0.35$ and $\eta = 0.05$.}
\end{figure}

For $\rho_g = 0.25$, and for $\rho_g=0.30$ in the low-noise regime, the prevailing velocity in the bands is $\mathbf{v}_1$. For the rest of the noises and densities analyzed, we see $\mathbf{v}_3$ emerge as the majority velocity within the condensed structures in the ordered phase. The final configurations encountered are illustrated in Figure~\ref{fig:TJIasy_Finals}.

\subsection{\label{subsec:TJII}Type-II Traffic Jam}

For the condensed structure denoted type-II traffic jam, a dense horizontal band was constructed and divided equally into two bands (see Fig.~\ref{fig:InitialConditions}(d)): at the top, the particles have velocity $\mathbf{v}_3$, and therefore tend to move down; they are then blocked by the particles at the bottom that are trying to go up since their velocities are $\mathbf{v}_2$. We consider densities $\rho_g = 0.32\ \mathrm{and}\ 0.48$. For $\rho_g = 0.48$, we have $\rho_s=1.0$ for the occupation probability within the band. However, for $\rho_g = 0.32$ this probability is $\rho_s = 0.7$, therefore the configuration does not strictly follow the definition for condensed structures and can be seen as a configuration to study the \textit{formation} of the TJ-II pattern.

\begin{figure}[!htp]
	\center
	\includegraphics[width=\linewidth]{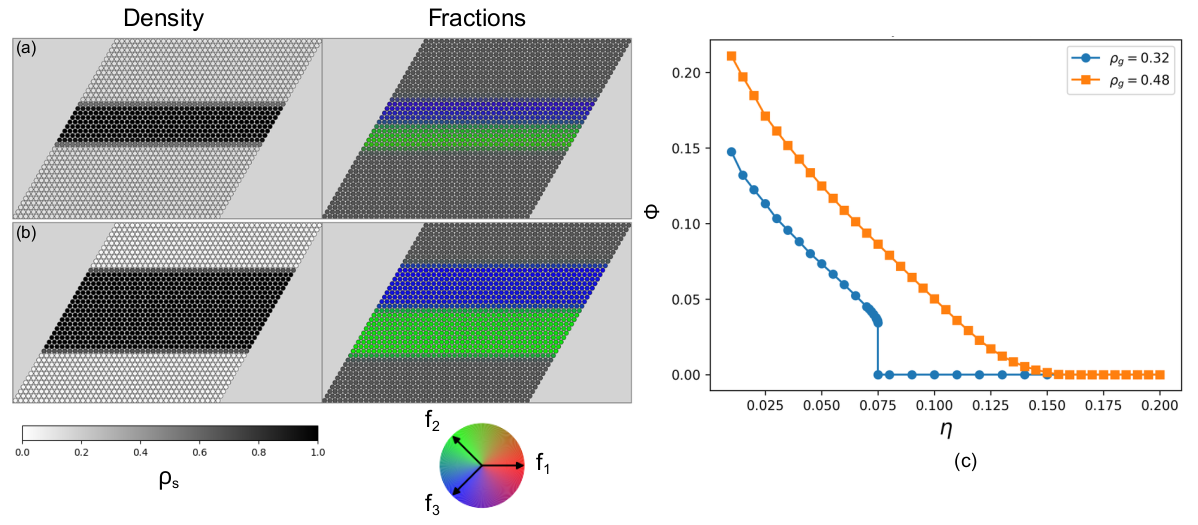}
	\caption{\label{fig:TJII_Finals} Final configurations for type-II traffic jam studies: (a) For $\rho_g = 0.32$ and $\eta=0.05$, the final configurations is a TJ-II with a band smaller than the initial one, but more populated. (b) The final configuration for $\rho_g = 0.48$ with $\eta=0.01$ is a TJ-II similar to the one constructed for the initial condition. (c) Order parameter as a function of noise for TJ-II, for $\rho_g = 0.32$ the transition to the disordered phase is discontinuous, while for $\rho_g=0.48$ the system goes to the disordered state continuously.}
\end{figure}

The transition to the disordered state is the only transition observed for this case; the steady state for $\eta < \eta_c$ is always a TJ-II. For $\rho_g = 0.32$, the initial dense region shrinks and becomes more populated in the early stages of the evolution, but the velocity distribution remains the same, as shown in Figure~\ref{fig:TJII_Finals}(a). For $\rho_g=0.48$, the decrease in band size occurs with increasing noise; for low noise, the initial and final configurations are very similar, as can be seen in Fig.~\ref{fig:TJII_Finals}(b). Figure~\ref{fig:TJII_Finals}(c) is consistent with the results shown in Figures~\ref{fig:IB_ordpar}(b)-(c) for IC III, where the TJ-II is formed. Low-density systems experience a discontinuous transition to disorder, while higher densities exhibit a continuous phase transition.

The results presented here correspond to configurations in which the initial dense band occupies half of the total lattice area. However, this structure has also been investigated for different band sizes, other densities, and even the case of homogeneous density as initial condition. In all these cases, the same qualitative behavior is observed, with the TJ-II pattern persisting for $\eta < \eta_c$.

\subsection{Random Initial Condition}

Studies of active matter frequently employ random initial conditions, which means that the system is essentially disordered. However, in MFT, a disordered state will never order, since there are no fluctuations. That is why we need to specify an initial configuration to analyze the condensed structures: the variations in density and fractions of particles in each state at different sites provide a perturbation that allows the system to evolve to states beyond the initial one. 

\begin{figure}[!htp]
	\center
	\includegraphics[width=0.85\linewidth]{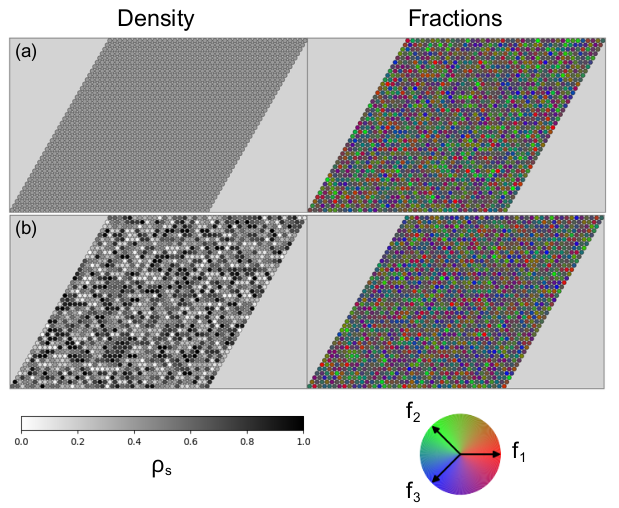}
	\caption{\label{fig:Rand_Initial} Random initial condition with the fractions of particles in each state chosen randomly on a lattice with (a) homogeneous density $\rho_g=0.5$, and (b) random density.}
\end{figure}

In addition to the condensed structures, we introduce two new initial conditions: (i) random fluctuations in the state fractions for each site $f_{i,s}$ with a homogeneous density distribution, and (ii) random fluctuations in the fractions and in the local density $\rho_s$. Both cases are shown in Figure~\ref{fig:Rand_Initial}.

\begin{figure}[!htp]
	\center
	\includegraphics[width=0.65\linewidth]{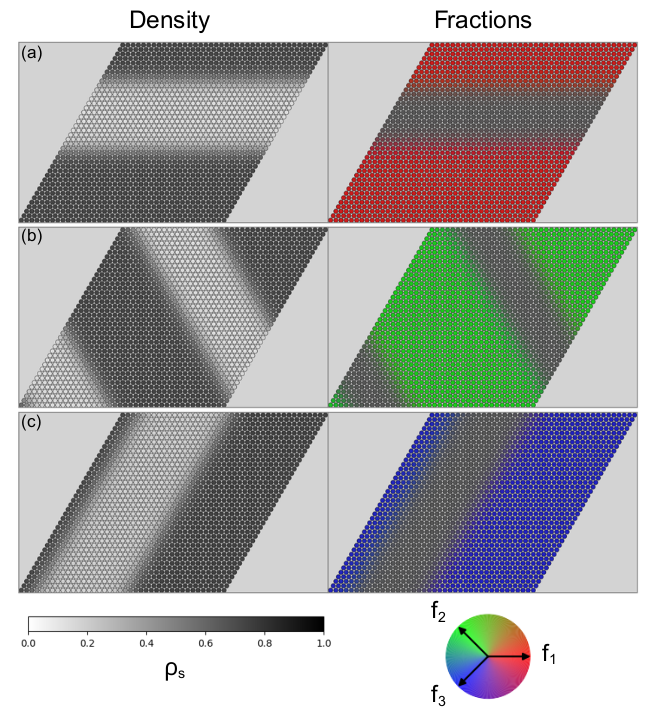}
	\caption{\label{fig:Rand_Finals} The final configuration for the random initial condition with fluctuations only in the fractions of particles with $\rho_g=0.5$ is an immobile band: (a) Emergence of an IB with majority $\mathbf{v}_1$ for $\eta = 0.02$. (b) Another realization with $\eta=0.02$ but now with two IBs with $\mathbf{v}_2$ as the majority. (c) An IB dominated by $\mathbf{v}_3$ for $\eta=0.03$.}
\end{figure}

For case (i), the results show that for $\eta < 0.05$ the system always evolves to form immobile bands oriented along $\mathbf{v}_1$, $\mathbf{v}_2$ or $\mathbf{v}_3$\cite{supp}, as shown in Figure~\ref{fig:Rand_Finals}. For $\eta \ge 0.05$, the noise overrides the alignment interaction, and the system rapidly reaches a homogeneous disordered state.

The majority orientation in the final configuration does not necessarily follow the majority initially defined by the random distribution nor is it a consequence of the noise intensity, since different final states are observed for the same noise value in different realizations of ICs with the same statistical properties.  

\begin{figure}[!htp]
	\center
	\includegraphics[width=0.65\linewidth]{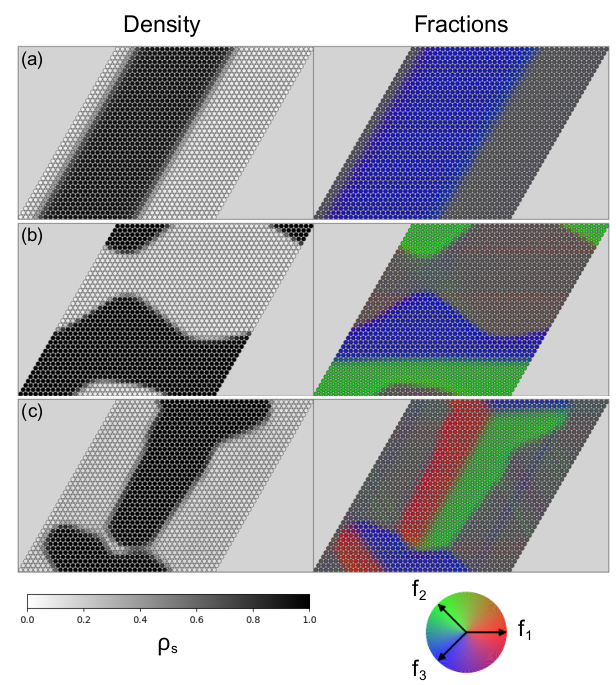}
	\caption{\label{fig:Rand_Finals2} Final configuration for the random $f_{i,s}$ and $\rho_s$ initial condition: (a) Emergence of an IB with majority $\mathbf{v}_2$ for $\rho_g = 0.498$ and $\eta = 0.02$. (b) Emergence of a TJ-II with velocities $\mathbf{v}_2$ and $\mathbf{v}_3$ for $\rho_g = 0.505$ and $\eta = 0.03$. (c) Emergence of a TJ-I for $\rho_g = 0.499$ $\eta=0.05$.}
\end{figure}

For initial condition (ii), we observe the creation of immobile bands and both types of traffic jams. For lower noise intensities, the steady state consists of an IB, as illustred in Fig.~\ref{fig:Rand_Finals2}(a). As $\eta$ increases, type-II traffic jams (Fig.~\ref{fig:Rand_Finals2}(b)) and type-I traffic jams (Fig.~\ref{fig:Rand_Finals2}(c)) emerge. It is worth noting that the global density $\rho_g$ is defined by the random distribution of the occupancy probability at each site, under the sole restriction that these probabilities are bounded between $[0, 1]$. The global average density of the system naturally fluctuates around the central value of $0.5$.

The results obtained for this IC differ significantly from those observed with ordered initial structures. Although the formation of bands and traffic jams persists, these structures emerge with distinct morphologies and dynamics. The TJ-II, for example, is a static structure when it originates from a condensed configuration; however, under random initial conditions, the structure propagates throughout the lattice as a wave \cite{supp}. In turn, the TJ-I extends across the lattice and maintains its stability even without equivalence between the state fractions.

%-------------------------------------------------------------------------------------------------%

\section{\label{sec:comp} Comparison with Simulation}

We now proceed to a comparison between the results of this study with those reported in \cite{rosembach2024} for the 3-SALGM with 19 neighbors. In that work, three different initial conditions were considered: random initial conditions, an immobile band, and a transverse band. The reported results indicate that the immobile band is the preferred condensed structure in the stationary state, as it either emerges or remains stable over a wide range of $\eta$ values across different densities. Other patterns typically appear in a small region of parameter space, emerging in the critical region.

The transverse band is characterized by an ordered region that extends perpendicular to the direction of motion. This case is not explored in the present work, as preliminary analyses of a transverse band as an initial condition indicated that the system evolves toward a homogeneous state in which the propagation orientation of the band dominates the entire lattice. Therefore, coexistence between disorder and condensed structures was not observed. For an initial configuration homogeneous in both density and fractions, homogeneity is preserved during the evolution, while the degree of order decreases with increasing noise intensity until a discontinuous phase transition drives the system to the disordered state \cite{dias2022}.

The behavior associated with the immobile band as initial condition is qualitatively similar in simulation and MFT. The IB structure persists as $\eta$ increases, but eventually gives way to other patterns. In the simulations, mobile bands and traffic jams were reported, whereas in the MFT framework only the TJ-II structure emerges.

In the case of random initial conditions, simulations reveal the emergence of immobile bands,  mobile bands, and traffic jams, indicating that different realizations may lead to distinct structures under the same control parameters. In the mean-field approximation, we observed the emergence of immobile bands and traffic jams for various noise values, and the pattern of these structures can differ for different realizations.

It is important to note that the random IC in \cite{rosembach2024} corresponds to the fully disordered state, equivalent to the disordered aggregate state observed at high noise. In contrast, in the mean-field description, what we denote as a random initial condition consists of a random assignment of $f_{i,s}$ and/or $\rho_s$ values, which implies that each site has a locally preferred orientation, whose contribution to alignment and movement dynamics is weighted by its occupancy probability. The disordered state in the MFT framework is instead the homogeneous configuration with $\rho_s = \rho_g$ and $f_{i,s} = 1/3$, $\forall s$, which is a stable solution of Eq.~(\ref{eq:eq_motion}). Escaping this state requires the introduction of fluctuations in the local fractions and/or in the density, thereby driving the system away from the homogeneous disordered state. Despite these differences, we have similar results for simulations and MFT.

While traffic jams do appear as steady-state configurations in simulations, they are restricted to a narrow region of the $\rho$-$\eta$ plane, between that corresponding to an IB and the transition to disorder. In contrast, the MFT yields TJ structures over a significantly larger region of the parameter space when starting from MB and asymmetric TJ-I initial conditions, though this region remains similarly constrained between IBs and disordered final states. Meanwhile, symmetric TJ-I and TJ-II configurations retain their initial structure up to the phase transition to the disordered state.

\subsection{\label{subsec:IBcomparison}Immobile Band: Details on Phase Transitions}

The results reported in Sec.~\ref{subsec:IB} show that when the lattice is initialized with an immobile band, a transition to a TJ-II is observed only at high densities; at low densities, the system undergoes a direct transition to the disordered state. This raised the question of whether this behavior reflects a limitation of the mean-field approach or whether a direct transition indeed occurs when an IB is introduced into a low-density lattice.

To address this question, simulations of the particle model described in Sec.~\ref{sec:model} were performed using an initial configuration consisting of an immobile band with velocity $\mathbf{v}_1$. The simulations were carried out for a lattice of size $L = 257$, using $10^7$ Monte Carlo steps. In each step, $N$ updates of randomly selected particles were attempted, where $N$ denotes the number of particles in the system.

\begin{figure}[!htp]
    \centering
    \includegraphics[width=0.75\linewidth]{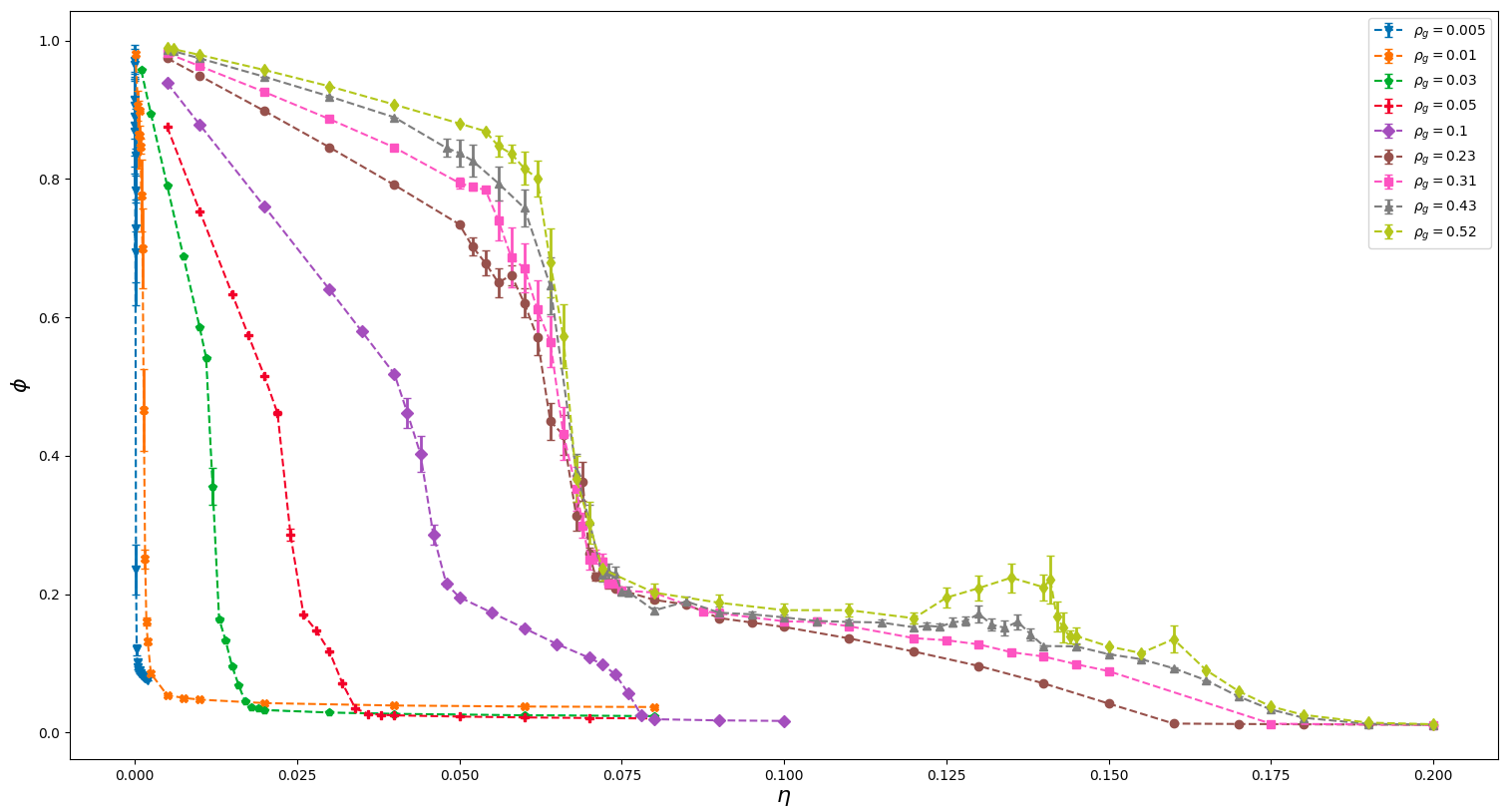}
    \caption{\label{fig:sim-IB}Simulation results for immobile band initial conditions on a lattice of size $L = 257$: for all densities analyzed the IB persists for low noise and exhibits a transition to a TJ-I configuration with increasing $\eta$, and attains a disordered state for further increase in $\eta$. Error bars that are not visible are smaller than the symbols.}
\end{figure}

Figure~\ref{fig:sim-IB} shows the order parameter $\phi = 1/N\ \left|\sum_{i=1}^N \mathbf{v}_i \right|$ observed in simulation as a function of noise for different densities. As in the MFT, the IB remains stable for low noise. However, in simulations the transition to the traffic jam occurs even at very low density, for example $\rho = 0.05$. This transition is actually to a type-I traffic jam; that is, starting from an IC with a dense band with $\mathbf{v}_1$ as the majority velocity, the system evolves to a final configuration with a dense region with three mutually blocking velocities. 

The simulations reveal nonmonotonic behavior of the fourth-order Binder cumulant with increasing noise, indicating a discontinuous order-disorder transition for densities below complete filling. The valleys, where the cumulant is typically negative, become less pronounced as the density increases. For full occupancy ($\rho = 1.0$), however, the Binder cumulant indicates a continuous phase transition, similar to that reported in the model of Venzel \textit{et al.}~\cite{rosembach2024}; see \cite{supp} for further details. Furthermore, simulations of the particle model were not able to determine the nature of the transition in cases where the system evolves between ordered states.

For densities $\rho = 0.43 \ \mathrm{and} \ 0.52$, a second phase transition occurs in which a type-II traffic jam emerges; this is evident in Fig.~\ref{fig:sim-IB} for $\eta \simeq 0.125$, where there is an increase in the order parameter. This transition is also observed in MFT. These results suggest that the mean-field approach cannot generate a TJ-I from a dense band; consequently, at lower densities the MFT predicts a direct transition to the disordered state. 

Revisiting the results for the asymmetric TJ-I in Sec.~\ref{subsec:TJIasym}, it appears that the TJ-I structures cannot arise from an asymmetric IC. Since the IB is also asymmetric, the formation of a TJ-I is not expected in this framework. Thus, transitions to TJ-I observed in simulations are not reproduced, whereas TJ-II structures are easier to obtain because the dense band persists while the majority velocities rearrange to form the jammed region.

In summary, the MFT consistently reproduces the main condensed phases and order-disorder transitions identified in simulations. The observed differences stem primarly from distinct initial condition implemetantions and inherent limits of the mean-field approximation.

\section{\label{sec:conclusion}Discussion and Conclusion}

\begin{figure}[!htp]
	\center
	\includegraphics[width=\linewidth]{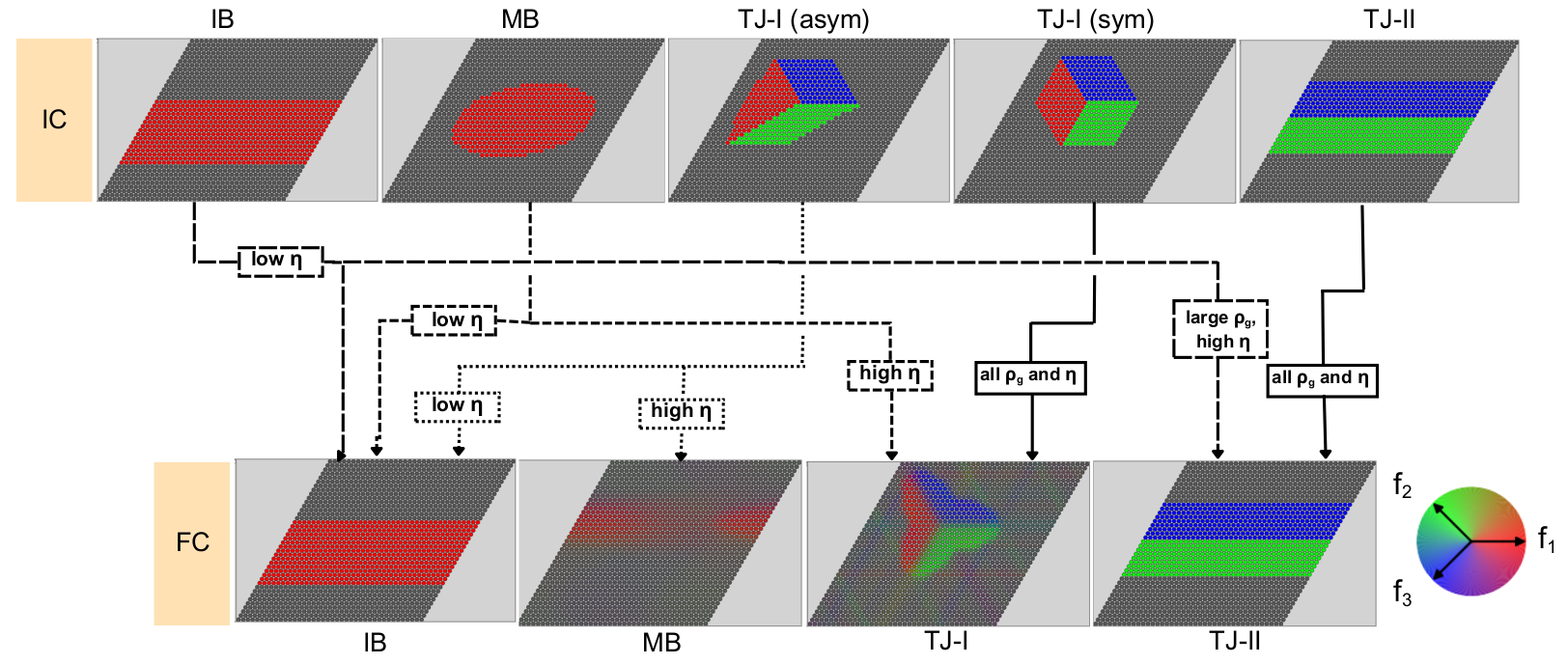}
	\caption{\label{fig:ICtoFC} The initial configurations (IC) for the condensed structures and the final configurations (FC) that each IC can reach for $\eta < \eta_c$.}
\end{figure}

We analyze the three-state active lattice gas model with excluded volume using a mean-field approximation including spatial structure, focusing on the formation and stability of the condensed structures characteristic of the model. To this end, we initialize the lattice in a configuration similar to the condensed structure under study and observe how it evolves for different values of density and noise. Figure~\ref{fig:ICtoFC} summarizes the ordered final configurations obtained from each initial configuration constructed. Videos illustrating the time evolution for the different initial conditions are provided in the Supplementary Material \cite{supp}.

Initializing the system with an immobile band, we observe two condensed structures in the steady state: the persistence of IB and the emergence of TJ-II. Upon comparison with simulation results, we encountered a limitation: when using the mean-field approximation for the 3-SALGM, we miss the emergence of a TJ-I. Thus, we have a direct transition to the disordered state when the density is low, while for higher densities, a transition to TJ-II occurs in agreement with the particle model simulations. However, both MFT and simulation results indicate the existence of a parameter region in which the immobile band is a stable structure, namely the low-noise regime. Moreover, the higher the density, the larger the stability region of the IB. In addition to identifying the conditions under which the immobile band persists, we also observe its formation from initial conditions with a mobile band, an asymmetric type-I traffic jam, as well as from random initial conditions.

The mobile band of Fig.~\ref{fig:InitialConditions}(b) proved to be an unstable structure, rapidly disappearing under all analyzed density and noise values. Initially, the MB expands to form an IB; then a different dynamics takes place inside the dense region, where particles reorient toward the minority velocities, leading to the formation of a type-I traffic jam. Although the constructed MB is not stable, a mobile band does emerge as a steady state for certain parameters of the asymmetric TJ-I initial condition. Comparing Figures~\ref{fig:InitialConditions}(b) and \ref{fig:TJIasy_Finals}(b)-(d), we see that these mobile bands differ significantly: the constructed one is approximately elliptical and occupies a large region in the direction perpendicular to the band orientation, whereas the stable ones are narrower, and are more elongated along the band orientation. This suggests that the model has a preferred form of organization for the condensed structures and points to a new configuration that may warrant further investigation.

Traffic jams are the most stable structures: both TJ-I (symmetric) and TJ-II persist for all the conditions studied, only disappearing when the system reaches the critical noise value, at which point the transition to the disordered state occurs. In the case of the TJ-I, although the system is initialized with a dense hexagonal structure, a rearrangement of the density and fraction distribution occurs, forming a clover-shaped structure identical to that observed in the transition from MB to TJ-I. This again points to an intrinsic structural arrangement favored by the dynamics. The mean-field approximation indicates that TJ-I can arise from an initial MB, random intial conditions and from a disordered IC with density fluctuations \cite{supp}, whereas TJ-II can be formed from an initial IB pattern and also from random initial conditions.

Type I traffic jams are of particular interest because they exhibit a stable structure characterized by the coexistence of all permitted velocities. Similar behavior is reported in the model with excluded volume and $q=4$ orientations, in which a 4-fold symmetric pattern emerges in accordance with the discretization \cite{peruani2011, karmakar2022}. Our understanding to date indicates that these \textit{traffic jams} are inherent in active models with restricted occupancy and $q$ discrete states, and that the observed symmetry depends directly on $q$. Although our model is restricted to $q=3$ for simplicity, we expect that extending the formalism to a larger number of states (such as $q=6$) will result in the emergence of traffic jams with $q$-fold symmetry.

The alternative configuration of the TJ-I with an asymmetric structure, in turn, always gives rise to bands. We observe the emergence of IB or MB, with velocities $\mathbf{v}_1$ or $\mathbf{v}_3$, depending on the initial conditions. The difference in results for the two types of traffic jams suggests that the initial spatial structure directly affects the final states. Specifically, a comparison between symmetric and asymetric TJ-I configurations at fixed global density ($\rho_g = 0.30$) reveals completely different final states for identical density and noise intensity. This behavior is an evidence of multistabilty governed by the initial spatial configuration.

Overall, the results demonstrate the combined effect of alignment, noise, and excluded volume on the stabilization and formation of condensed structures. The response of the system to different initial conditions reveals that each converges to a distinct steady state under the influence of these three interactions. In general, the fractions of particles in each state $f_i$ inside an ordered region are lower than their initial values -a divergence that increases monotonically as noise increases. This behavior arises from the stochasticity introduced by noise, which destabilizes the alignment mechanism and reduces the probability of particles transitioning to the majority state. Under low-noise conditions, the alignment interaction predominates, preserving order. However, the progressive increase in noise intensity for the same initial configuration favors the growth of minority fractions. This process explains, for example, the formation of traffic jams from IB and MB: when noise is sufficiently high, subpopulations begin to form within the band whose most probable state converges towards the minority orientation. Combined with the restriction imposed by excluded volume - which prevents particles from leaving high-density regions- this population stabilizes in the form of jammed structures with two or three orientations. Traffic jams are also subject to this effect, but because of the coexistence of different orientations, the increase in the probabilities of minority states induced by noise drives the system directly into the disordered phase, rather than causing a reversal of the majority state.

Starting from random initial conditions, the final condensed structures are immobile bands and traffic jams of both kinds. Rather than being artifacts of ordered initial conditions, these results demonstrate the capability of the mean-field formalism to reproduce the characteristic patterns observed in self-propelled models with excluded-volume interactions. The creation and maintenance of these structures are fundamentally driven by spatial variations that generate state density fluxes between adjacent sites, as modeled by Eq.~(\ref{eq:density_transfer}). Indeed, if the system is initialized in a spatially homogeneous configuration with a small perturbation in the state fractions, the dominance of the initial majority state is observed for $\eta < \eta_c$, in accordance with the alignment interaction, but without segregation into dense regions \cite{dias2022}. In this situation, the rates at which a site receives and donates density are equal, yielding a net zero current throughout the lattice and preserving spatial homogeneity. Thus, for condensed structures to emerge, it is imperative that the initial conditions for $\rho_s$ and/or $f_{i,s}$ exhibit spatial variation. Furthermore, our results indicate that the set of equations developed for local state density variations, together with the density flux between sites, consistently describes the instabilities associated with the emergence of condensed structures. A formal stability analyses of the equations will be the focus of future work.

The hexagonal topology of the lattice sites bears similarities to the spatial arrangements of epithelial tissue layers. From a phenomenological perspective, the proposed model provides a basis for modeling this type of biological system; however, it would be essential to incorporate a mechanism for cell wall deformation in response to local density flows. This deformability would then alter the coordination number of the cell, directly affecting the alignment interactions and the global order of the system \cite{nejad2025, nejad2025-2, Happel2025, Maria2023}.

The results obtained from the mean-field approximation and from simulations of the particle model offer a starting point for further investigation and support the consistency of the proposed framework. As expected, this type of approximation presents some limitations: the inability to study very large systems, along with the simplifications introduced by the mean-field approach, prevent a precise determination of the nature of the observed phase transition and may even mask some of them, as seen in this work. Nevertheless, MFT offers a qualitative description of phase transitions, and within the context of active matter models with excluded volume, reproduces the expected patterns and collective behaviors. It also reveals transitions between ordered states that were not initially anticipated.  Furthermore, the approach allowed us to investigate the stability conditions and formation of condensed structures in the three-state active lattice gas, showing, in most cases, qualitative agreement with simulation results.

%-------------------------------------------------------------------------------------------------%
\section*{Data Availability Statement}
The data that support the findings of this article are publicly available \cite{zenodo2}. The supplementary videos for this article are available in \cite{zenodo}. The source code used to generate the data is available upon reasonable request.

%-------------------------------------------------------------------------------------------------%
\section*{Acknowledgments}
This work was supported by the Brazilian agencies CNPq and FAPEMIG. The computations and simulations were performed using the cluster of \textit{Departamento de Física, ICEx, Universidade Federal de Minas Gerais} (UFMG), as well as the cluster of the \textit{Instituto Nacional de Ciência e Tecnologia de Sistemas Complexos} (INCT-SC). We are grateful to Prof.\ Gerald Weber and to INCT-SC for their support, which was essential for the functioning of these resources.

%-------------------------------------------------------------------------------------------------%

\appendix

\section{Density Transfer Process}
To begin, we analyze the transfer of a single species $i$, from site $s$ to a nearest neighbor $s'$ accessible to species $i$. Consider the transfer of density $i$ (with $i = 1,\ 2\ \mathrm{or}\ 3$) from site $s$ with type-$i$ density $g(t) = \rho_s$. Its neighbor, site $s'$, has a density of $r(t) = \rho_{s'}$. The current density from site $s$ to $s'$, denoted $J_{\rho} (t)$, follows the relation:
\begin{equation}
    \frac{d}{dt}r(t) = - \frac{d}{dt}g(t) = J_{\rho} (t),
\end{equation}
where
\begin{equation}
    J_{\rho}(t) = v_0 g(t) \left[1-r(t) \right],
\end{equation}
with $v_0$ the velocity magnitude and $\left[1-r(t) \right]$  the density that site $s'$ can accommodate. Letting $u(t) = 1 - r(t)$, we have:
\begin{equation}
    \frac{d}{dt}u(t) = \frac{d}{dt}g(t) = - J_{\rho}(t),
\end{equation}
where
\begin{equation}
    J_{\rho} = v_0 g(t) u(t).
\end{equation}

Let $g_0$ and $u_0$ denote the densities at some time $t_0$. The source site can transfer an amount $\Delta g$ of density to the receiver. The time $dt$ required for the source density to change from $g_0$ to $g_0 - \Delta g \ge 0$ is,
\begin{equation}
    dt = -\frac{dg}{J_{\rho}} = - \frac{1}{v_0 g_0 u_0}.
\end{equation}
Simultaneously, the receiving site must increase its density by $\Delta g$, leading to the relation:
\begin{equation}
    \int_{0}^{\Delta t} dt = -\frac{1}{v_0} \int_{0}^{\Delta g} \frac{dy}{(g_0 -y)(u_0 - y)}.
\end{equation}

Two scenarios may arise during this process: the source can send an amount equal to what the receiver can accept, $g_0 = u_0$, or these densities can differ, $g_0 \ne u_0$. Both cases must be addressed when solving the integral.

\begin{itemize}
    
  %  \begin{widetext}
    \item If $g_0 = u_0$:
         \begin{equation*}
        \begin{split}
            \Delta t = -\frac{1}{v_0} &\int_{0}^{\Delta g} \frac{dy}{\left(g_0 - y \right)^2} = \frac{1}{v_0}\frac{1}{\left(y - g_0 \right) }\Big |_{y=0}^{y=\Delta g} \\
            \Delta t &= \frac{1}{v_0}\frac{\Delta g}{g_0\left(\Delta g - g_0\right)}.
        \end{split}
    \end{equation*}
        Thus, we have
    \begin{equation}
        \label{eq:flux1}
        \Delta g = \frac{v_0 \Delta t g_0^2}{v_0 \Delta t g_0 - 1}.
    \end{equation}
  %  \end{widetext}

    \begin{widetext}
    \item If $g_0 \ne u_0$:
        \begin{equation*}
        \begin{split}
            \Delta t = -\frac{1}{v_0}\int_{0}^{\Delta g} \frac{dy}{\left(g_0 -y \right)\left(u_0 - y\right)} &= \frac{1}{v_0 \left(g-0 - u_0 \right)} \left[\ln\left(y-u_0 \right) - \ln\left(y - g_0 \right) \right]\Big |_{y=0}^{y=\Delta g} \\
            \Delta t &= \frac{1}{v_0 \left(g_0 - u_0 \right)}\ln \left(\frac{\Delta g/u_0 - 1}{\Delta g/g_0 -1} \right).
        \end{split}
    \end{equation*}
    Consequently, we find:
    \begin{equation}
        \label{eq:flux2}
        \Delta g = - \frac{g_0\left[e^{v_0\Delta t(g_0 - u_0)} - 1\right]}{e^{v_0 \Delta t (g_0-u_0)} - g_0/u_0}
    \end{equation}
    \end{widetext}
    
\end{itemize}

\section{Arrangement of Condensed Structures}

By definition, the condensed structures are ordered patterns with low-mobility due to the restriction imposed by excluded volume. In the particle model framework, dense structures like bands and traffic jams do not exhibit particle motion in their interior due to the excluded volume; transport occurs exclusively at the boundaries of the structure, where the rearrangement of local orientations can cause particle evaporation.

In MFT the dynamics of movement is more subtle, since the occupancy is treated as a continuous variable. The evolution of the system under the influence of alignment and noise can result in a decrease in local density in regions of high occupancy, a phenomenon driven by the density difference at the borders. Equations \ref{eq:flux1} and \ref{eq:flux2} allow density flows whenever $0 < \rho_s < 1$, so a current is established within the dense structure. However, this should not be understood as the movement of a particle: when $\rho_s = 1$ site $s$ is occupied with probability equal to one; consequently, any flux that tends to increase this density is blocked, as it would imply more than one particle at the site.

  \begin{figure}[!htp]
    \centering
    \includegraphics[width=0.85\linewidth]{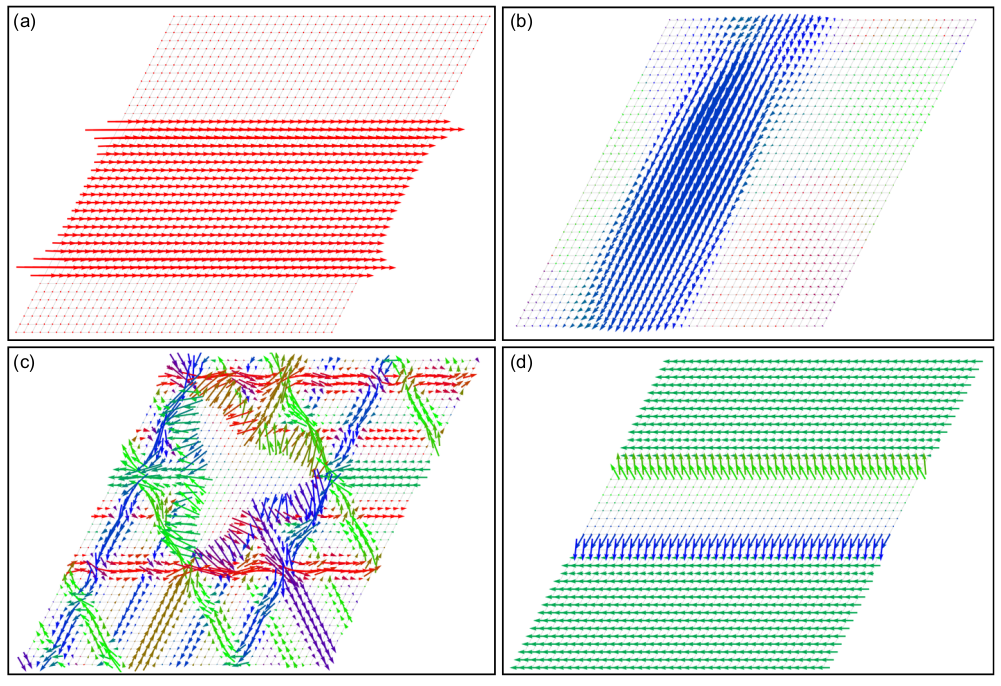}
    \caption{\label{fig:currents}Density currents for: (a) Immobile Band with $\rho_g = 0.43$ and $\eta = 0.01$. (b) Mobile Band with $\rho_g = 0.30$ and $\eta = 0.033$. (c) Type-I Traffic jam with $\rho_g = 0.30$ and $\eta = 0.05$. (d) Type-II Traffic Jam with $\rho_g = 0.32$ and $\eta = 0.01$. The arrows indicate the direction of the liquid current for each site. As usual, red corresponds to $\mathbf{v}_1$, green to $\mathbf{v}_2$, and blue to $\mathbf{v}_3$.}
\end{figure}

In Figure~\ref{fig:currents}, we present the net current for each site $\mathbf{J}_s$ for the density associated with the condensed structures. The figures highlight the counterintuitive aspect of the dynamics of motion in mean-field theory. The bands are illustrated by Fig.~\ref{fig:currents}(a)-(b), indicating a more intense flow within the dense structures. This occurs because, outside this region, the density is extremely low, causing $\mathbf{J}_s \to 0$. In contrast, a higher flux density is allowed inside the band, where $\rho \to 1.0$ and gradually decreases at the borders. For the MB, the region with the highest current represents the translation of the band itself along the lattice.
 
 In contrast, traffic jams exhibit small particle currents in their interiors, evidence of the congestion caused by the collision of opposing velocities. As expected, a stronger flow is observed at the boundaries. In the region of low occupancy, random flows are observed for TJ-I in Fig.~\ref{fig:currents}(c), while for the TJ-II in Fig.~\ref{fig:currents}(d) the flow is oriented in the direction resulting from the combination of velocities $\mathbf{v}_2$ and $\mathbf{v}_3$.

The density current at site $s$ is given by
\begin{equation}
    \label{eq:current_tot}
    \mathbf{J}_s = \sqrt{\left(J_{x,s} \right)^2 + \left(J_{y,s} \right)^2}.
\end{equation}
Where the components $x$ and $y$ of the current are defined by:
\begin{equation}
    \begin{split}
        J_x = v_{1,x} \Delta g_{v_1} + v_{2,x} \Delta g_{v_2} + v_{3,x} \Delta g_{v_3},\\
        J_y = v_{1,y} \Delta g_{v_1} + v_{2,y} \Delta g_{v_2} + v_{3,y} \Delta g_{v_3}.
    \end{split}
\end{equation}
Here, $v_{i,x}$ and $v_{i,y}$ represent, respectivelly, the $x$ and $y$ components of the velocity vector, as per Eq.~(\ref{eq:velocities}); while $\Delta g_{v_i}$ denotes the amount of density transferred by site $s$ along the direction given by $\mathbf{v}_i$.

\nocite{pottsmodel1982}
\bibliography{MeanField}

\end{document}